%% file: ms.tex
\documentclass[fleqn,usenatbib]{mnras}
\usepackage{newtxtext,newtxmath}
\usepackage[T1]{fontenc}
\usepackage{ae,aecompl}
\usepackage{graphicx}	 
\usepackage{epstopdf}   
\usepackage[shortcuts]{extdash}    
\usepackage{threeparttable}
\usepackage[dvipsnames]{xcolor}
\DeclareGraphicsExtensions{.png,.pdf}
\newcommand\eso        {ESO~137\=/002}
\newcommand\hst        {{\em HST}\/}
\newcommand\apex       {{\em APEX}\/}

\newcommand{\hii}{H\,{\sc ii}}
\newcommand{\RNum}[1]{\uppercase\expandafter{\romannumeral #1\relax}}
\newcommand{\NII}{\mbox{[N\,\textsc{ii}]}}

\title[A RPS galaxy with little SF in the tail]{\eso{}: a large spiral undergoing edge-on ram-pressure stripping with little star formation in the tail}

\author[S. Laudari et al.]{
Sunil Laudari$^{1}$\thanks{E-mail: ming.sun@uah.edu},
Pavel J\'{a}chym$^{2}$\thanks{E-mail: jachym@ig.cas.cz},
Ming Sun$^{1}$\thanks{E-mail: ming.sun@uah.edu},
Will Waldron$^{1}$,
Marios Chatzikos$^{3}$,
\newauthor{}
Jeffrey Kenney$^{4}$,
Rongxin Luo$^{1}$,
Paul Nulsen$^{5,6}$,
Craig Sarazin$^{7}$,
Fran\c{c}oise Combes$^{8}$,
\newauthor{}
Tim Edge$^{1}$,
Mark Voit$^{9}$,
Megan Donahue$^{9}$,
Luca Cortese$^{6}$
\\
\\
$^{1}$Department of Physics \& Astronomy, University of Alabama in Huntsville, 301 Sparkman Dr NW, Huntsville, AL 35899, USA\\
$^{2}$Astronomical Institute of the Czech Academy of Sciences, Bo\v{c}n\'{i} II 1401, 141 00, Prague, Czech Republic\\
$^{3}$Department of Physics \& Astronomy, University of Kentucky, Lexington, KY 40506, USA\\
$^{4}$Yale University Astronomy Department, P.O. Box 208101, New Haven, CT 06520-8101, USA\\
$^{5}$Center for Astrophysics \textbar{} Harvard \& Smithsonian, Cambridge, MA 02138, USA \\
$^{6}$International Centre for Radio Astronomy Research, University of Western Australia, 35 Stirling Hwy, Crawley, WA 6009, Australia\\
$^{7}$Department of Astronomy, University of Virginia, Charlottesville, VA 22904, USA\\
$^{8}$Observatoire de Paris, LERMA, PSL, CNRS, Sorbonne Univ. UPMC, and College de France, F-75014, Paris, France \\
$^{9}$Department of Physics and Astronomy, Michigan State University, East Lansing, MI 48824, USA\\
}

\date{Accepted XXX. Received YYY; in original form ZZZ}

\pubyear{20XX}

\begin{document}
\label{firstpage}
\pagerange{\pageref{firstpage}--\pageref{lastpage}}
\maketitle

\begin{abstract}
Ram pressure stripping (RPS) is an important mechanism for galaxy evolution.
In this work, we present results from \hst{} and \apex{} observations of one RPS galaxy, \eso{} in the closest rich cluster Abell~3627.
The galaxy is known to host prominent X-ray and H$\alpha$ tails.
The \hst{} data reveal significant features indicative of RPS in the galaxy, including asymmetric distribution of dust in the galaxy, dust filaments and dust clouds in ablation generally aligned with the direction of ram pressure, and young star clusters immediately upstream of the residual dust clouds that suggest star formation (SF) triggered by RPS.
The distribution of the molecular gas is asymmetric in the galaxy, with no CO upstream and abundant CO downstream and in the inner tail region. A total amount of $\sim 5.5 \times 10^{9}$ M$_\odot$ of molecular gas is detected in the galaxy and its tail.
On the other hand, we do not detect any active SF in the X-ray and H$\alpha$ tails of \eso{} with the \hst{} data and place a limit on the SF efficiency in the tail. 
Hence, if selected by SF behind the galaxy in the optical or UV (e.g., surveys like GASP or using the {\em Galex} data), \eso{} will not be considered a ``jellyfish'' galaxy. Thus, 
galaxies like \eso{} are important for our comprehensive understanding of RPS galaxies and the evolution of the stripped material.
\eso{} also presents a great example of an edge-on galaxy experiencing a nearly edge-on RPS wind.
\end{abstract}


\input{intro.tex}

\input{HST_observ.tex}

\input{APEX_observ.tex}

\input{photometry.tex}

\input{HST_src_pop.tex}

\input{CO_emission.tex}

\input{discuss.tex}

\input{conclude.tex}

\input{acknowledge.tex}

\appendix

\input{color_excess.tex}

\bibliographystyle{mnras}

\input{bibliography.bbl}
\bsp	
\label{lastpage}
\end{document}

%% file: intro.tex
\section{Introduction}\label{sec:intro}

Galaxy clusters are the largest gravitationally collapsed structures in the universe, filled with hot diffuse plasma with a typical temperature of 10$^{7} - 10^{8}$ K and electron number density of $10^{-4} - 10^{-1}$ cm$^{-3}$~\citep[e.g.,][]{2012ARA&A..50..353K}. They contain a large fraction of early-type galaxies mostly concentrated around the cluster core and have a relatively small fraction of spirals/irregulars~\citep[e.g.,][]{1980ApJ...236..351D}. Cluster galaxies are preferentially red with little cold gas and suppressed star formation~(SF) activity.
Conversely, field galaxies are mostly 
spirals that are preferentially blue, exhibit more cold gas, and have active SF. These characteristics suggest that galaxy evolution is strongly influenced by the environment~\citep{2006PASP..118..517B}. 
Galaxy clusters are cosmic labs for studying the effect of the environment on galaxy morphology and evolution.
Ram pressure stripping (RPS) \citep{1972ApJ...176....1G} is one of the most important mechanisms for evolution of gas-rich galaxies in clusters by 
removing hot and cold interstellar medium~(ISM) from the galaxy. As the galaxy moves through the intra-cluster medium~(ICM), SF can be suppressed as the ISM is lost. 
There can also be a momentarily initial starburst, as thermal instabilities and turbulent motions trigger the collapse of the molecular clouds before SF is eventually quenched~\citep[e.g.][]{2003ApJ...596L..13B}. The fate of the stripped ISM is also an interesting question. It eventually should get mixed with the ICM to enrich its metallicity~\citep[e.g.,][]{2008SSRv..134..363S}.
On the other hand, it is now known that some fraction of the stripped ISM can turn into stars in the galactic halo and the intra-cluster space,
even though the quenching happens within the galaxy \citep[e.g.,][]{2007ApJ...671..190S, 2007ApJ...660.1209Y,2010MNRAS.408.1417S,2010AJ....140.1814Y,2014ApJ...792...11J,2017ApJ...844...48P}.

Observational evidence of RPS has been revealed in many different bands. Stripped gaseous tails and star-forming clumps have been detected in UV and H$\alpha$~\citep[e.g.,][]{2001ApJ...563L..23G,2007ApJ...671..190S,2007MNRAS.376..157C,2007ApJ...660.1209Y,2008ApJ...688..918Y,2010MNRAS.408.1417S,2010AJ....140.1814Y,2010ApJ...716L..14H,2014ApJ...780..119K,2016MNRAS.455.2028F}.
\ion{H}{I} tails in some late-type galaxies in the Virgo cluster and other clusters have also been discovered~\citep[e.g.,][]{2005A&A...437L..19O, 2007ApJ...659L.115C}.
Stripped tails have also been revealed in X-ray~\citep[e.g.][]{2004ApJ...611..821W, 2005ApJ...621..718S,2006ApJ...637L..81S,2010ApJ...708..946S} and radio~\citep[e.g.,][]{1985ApJ...292..404G, 1987A&A...186L...1G,2020MNRAS.496.4654C}. \citet{2014ApJ...792...11J, 2017ApJ...839..114J, 2019ApJ...883..145J} detected a large amount of cold molecular gas, traced by the CO emission, in the stripped tails behind ESO~137-001 and D100, embedded in the hot ICM. 
Also, cold molecular gas has been detected in the tails of other galaxies undergoing RPS \citep{2015A&A...582A...6V,2018MNRAS.480.2508M, 2020ApJ...889....9M}.
Warm molecular H$_{2}$ gas clouds have also been revealed in the tails of ESO~137-001 and other RPS galaxies~\citep{2010ApJ...717..147S, 2014ApJ...796...89S}. With the detection of active SF in many RPS galaxies, 
these galaxies are often called `jellyfish galaxies'.

RPS has also been studied extensively using numerical simulations \citep[e.g.][]{1999MNRAS.308..947A,2000Sci...288.1617Q, 2001MNRAS.328..185S,2001ApJ...561..708V,2006MNRAS.369..567R,2007A&A...472....5J,2009A&A...499...87K,2012MNRAS.422.1609T,2014ApJ...784...75R}.
These simulations demonstrate important effects of RPS on galaxy evolution like disk truncation, SF quenching, central bulge build-up, formation of flocculent arms, the transformation of dwarf galaxies and long filamentary structures. The overall results of these simulations have brought to a more realistic comparison with the observation data.

\begin{figure*}
    \includegraphics[width=0.99\textwidth]{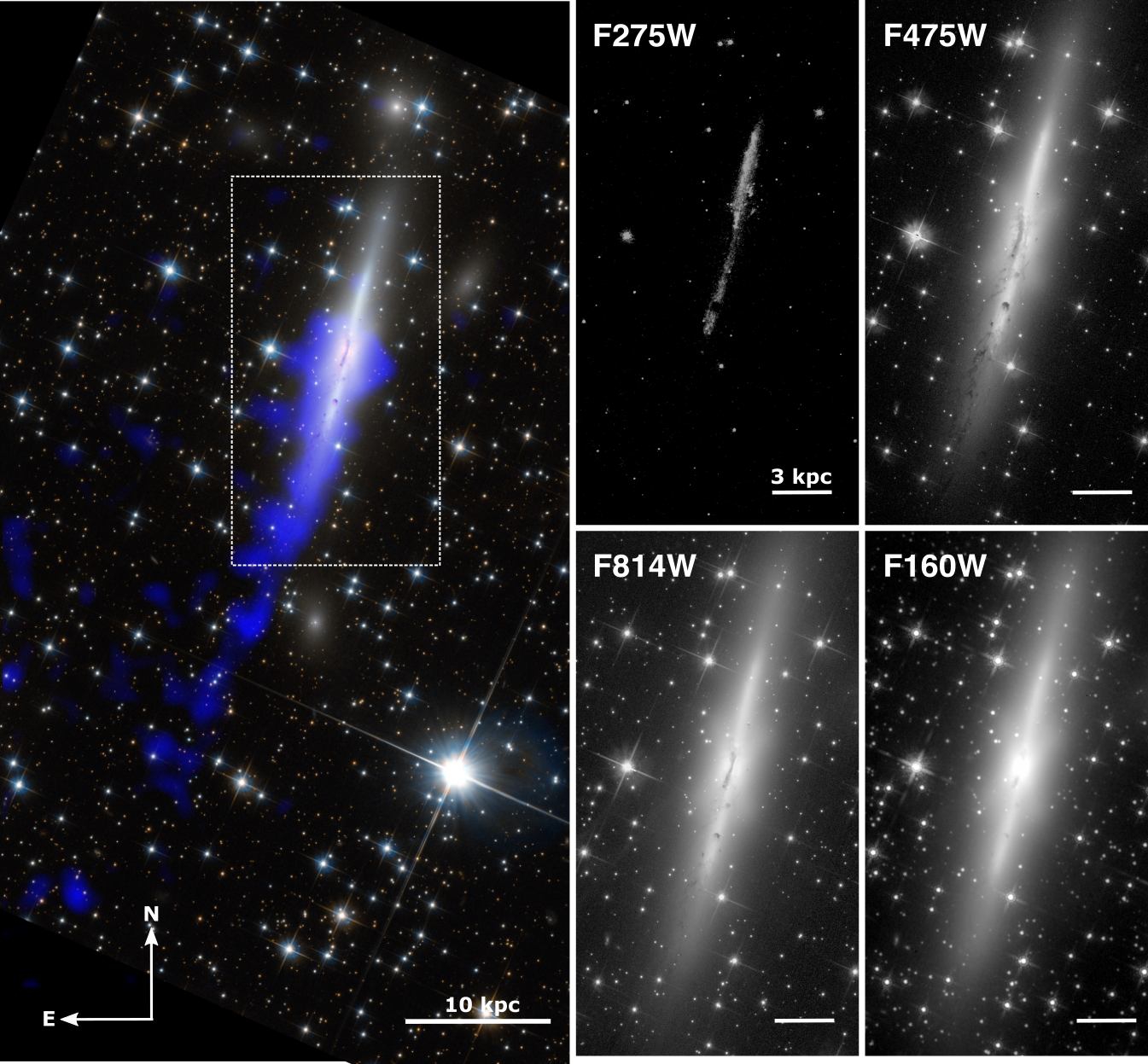}
    \vspace{-0.2cm}
    \caption{{\em Left}: Composite image of \eso{} with the \hst{} F475W/F814W data
    (Credit: ESA/Hubble \& NASA). We also overlay the 0.6 - 2 keV X-ray emission from {\em Chandra} in blue from \citet{2013ApJ...777..122Z} to show the X-ray tail. The galaxy region inside the white box in the dotted line is highlighted in the right panels. {\em Right}: A close-up view of \eso{} in the four \hst{} bands (see Table~\ref{tab:hstObs1} for the detail of these four bands). While the upstream of the galaxy is dust free, many dusty clouds and filaments are observed in the downstream. One can also observe the X-shaped bulge, especially on the red bands.}    
    \label{fig:composite}
\end{figure*}

In this paper, we focus on \eso{}~(Fig.~\ref{fig:composite}), which is a large and bright late-type galaxy in the closest rich cluster Abell~3627 and hosts a Seyfert2-like active galactic nucleus (AGN)~\citep{2010ApJ...708..946S,2013ApJ...777..122Z}. 
It is at a projected distance of only $\sim$ 110 kpc (or $\sim 0.06$ $r_{200}$) from the cluster's X-ray center and has a radial velocity of $\sim$ +820 km/s with respect to that of the cluster. 
Table \ref{tab:property} lists some properties of \eso{}. It is $\sim$ 5 times more massive than its neighbor RPS galaxy ESO~137-001 \citep{2010ApJ...708..946S} but with a similar SFR and a similar optical size.
While stripping in ESO~137-001 is close to face-on, \eso{} is experiencing a near edge-on stripping, which makes it an important comparison with ESO~137-001 for detailed studies on RPS and galaxy evolution. 
With an axis ratio of 0.3 from the {\em H}-band image and assuming a morphological type from Sa to Sb (Sab),
using the classical Hubble formula, the inclination angle, measured between the line-of-sight and the disk axis, is $\sim$ 88$^{\circ}$ - 78$^{\circ}$ (81$^{\circ}$) so we are viewing the galaxy nearly edge-on (an average inclination angle of 83$^{\circ}$ is then adopted in this work).
Downstream of \eso{}, there is a $\sim$ 40 kpc narrow X-ray tail~\citep{2010ApJ...708..946S,2013ApJ...777..122Z} with a constant width of $\sim$ 3 kpc~\citep{2013ApJ...777..122Z}.
\eso{} also features double H$\alpha$ tails extending $\sim$ 20 kpc from the nucleus without any \ion{H}{II} regions identified~\citep{2010ApJ...708..946S}. The secondary tail ($\sim$ 12 kpc) resides at a distance of $\sim$ 7.5 kpc from the nucleus at a $\sim$ 23$^\circ{}$ to the main tail.
The H$\alpha$ main tail is spatially coincident with the X-ray tail. The sharp H$\alpha$ leading edge is also at the same position of the X-ray leading edge. As emphasized in \cite{2010ApJ...708..946S} and \cite{2013ApJ...777..122Z}, the existing H$\alpha$ data of \eso{} are quite shallow and its X-ray data are also not deep for its proximity to the bright cluster core. The true extent of \eso{}'s X-ray and H$\alpha$ tails can be much longer than what is revealed from the current data.

\begin{table}
\begin{center}
    \caption{Properties of \eso{}}
    \label{tab:property}
    \begin{tabular}{c|c}
    Parameter                                & \eso{}\\
        \hline
        \hline
        Heliocentric velocity (\text{km/s})$^\mathrm{a}$    &  5691 (+820)   \\
        Distance (Mpc)$^\mathrm{b}$                         &  69.6    \\
        Offset (kpc)$^\mathrm{c}$                                       & 110 \\
        Position Angle                              & $\sim$ 13.1$^\circ{}$ \\
        Inclination                              & $\sim$ 83$^\circ{}$ \\
        $W1$ (Vega mag)$^\mathrm{d}$ & 10.21 \\
        $L_{\rm W1}$ ($10^{9} L_{\odot}$)$^\mathrm{d}$ & 3.39 \\
        $W1 - W4$ (Vega mag)$^\mathrm{d}$ &  4.25  \\
        $m_{\rm F160W}$ (AB mag)$^\mathrm{e}$ & 11.39 \\
        Half light semi-major axis (kpc)$^\mathrm{e}$   &  3.30 \\
        $M_{\star}$ ($10^{9} M_{\odot})^\mathrm{f}$         & 32-39  \\
        $M_{\rm mol}$ ($10^9 M_\odot$)$^\mathrm{g}$ & $\sim$ 5.5 \\
        $L_{\rm FIR}$ ($10^{10} L_{\odot})^\mathrm{h}$  & 1.22 \\ 
        SFR (M$_{\odot}$/yr)$^\mathrm{i}$                   & 0.94 \\
        Tail length (kpc)$^\mathrm{j}$                      & 40 (X-ray), $>$ 20 (H$\alpha$)\\
        \hline
    \end{tabular}
\end{center}
    Note: \\
   $^\mathrm{(a)}$ The heliocentric velocity from~\citet{2003AJ....126.2268W}. The velocity value in parentheses is the radial velocity relative to that of Abell~3627.\\
   $^\mathrm{(b)}$ For consistency, we use the cluster redshift ($z$ = 0.0163) and the luminosity distance used in \citet{2010ApJ...708..946S}. 1$''$ = 0.327 kpc.\\
    $^\mathrm{(c)}$ The projected offset of the galaxy from the X-ray center of A3627 \\
    $^\mathrm{(d)}$ The {\em WISE} 3.4 $\mu$m magnitude, luminosity and the {\em WISE} 3.4 $\mu$m - 22 $\mu$m color. The Galactic extinction was corrected with the relation from \cite{2005ApJ...619..931I}. \\
    $^\mathrm{(e)}$ The total magnitude and the half light semi-major axis at the F160W band. \\
    $^\mathrm{(f)}$ The total stellar mass estimated from \citet{2010ApJ...708..946S}.\\
    $^\mathrm{(g)}$ The total amount of molecular gas detected from \eso{} from this work \\ 
    $^\mathrm{(h)}$ The total FIR luminosity from the {\em Herschel} data (see Section~\ref{subsec:SF_gal})\\
    $^\mathrm{(i)}$ The average value from the first estimate (1.08) based on the {\em Galex} NUV flux density and the total FIR luminosity from {\em Herschel} with the relation from \citet{2011ApJ...741..124H}, and the second estimate (0.80) based on the {\em WISE} 22 $\mu$m flux density with the relation from \citet{2013ApJ...774...62L}. The Kroupa IMF is assumed.\\ 
    $^\mathrm{(j)}$ Tail length from~\citet{2013ApJ...777..122Z}
\end{table}

This paper presents new results on \eso{} from the {\em Hubble Space Telescope} (\hst{}) and the {\em Atacama Pathfinder EXperiment telescope}~(\apex{}). We focus on the main body and tail of \eso{} (regions of interests defined in Fig.~\ref{fig:region}). 
Sections \ref{sec:observation} and \ref{sec:CO_observation} give details about the observations and data reduction for \hst{} and \apex{}, respectively. Section \ref{sec:morph} presents the \hst{} results on \eso{} itself.
Section \ref{sec:hst_srcs} studies the \hst{} source population in \eso{}'s tail. 
Section \ref{sec:CO_emission} presents the CO emission in the galaxy and its tail. 
Section \ref{sec:discussion} is the discussion and we present the final conclusions in Section \ref{sec:conclusion}.

%% file: HST_observ.tex
\section{\hst{} Observations and Data Analysis}\label{sec:observation}
\eso{} was observed in the \hst{} program 12372 (PI:Sun), with two \hst{} imaging instruments --- Advance Camera for Surveys~(ACS) and --- Wide Field Camera 3~(WFC3). Table \ref{tab:hstObs1} lists details of the observations. The F475W and F814W high-throughput filters in the ACS/Wide Field Channel (WFC) allow the detection of faint features (e.g. faint star clusters in the tail) and a deep color map of the galaxy. The F275W filter in the WFC3/UVIS channel
adds important information on the SF in the last several $10^{7}$ year. On the other hand, the F160W filter in the WFC3/IR channel is the band least affected by the dust extinction. When combined, they can be used to constrain the stellar age of star clusters found in the tail and the galaxy.  The foreground Milky Way (MW) extinction is high in the direction of \eso{} as it lies near the Galactic plane, 1.096 mag for F275W, 0.653 mag for F475W, 0.305 mag for F814W, and 0.102 mag for F160W, with the extinction law and the dust map from \citet{1999PASP..111...63F} and \citet{ 2011ApJ...737..103S} respectively.
The 3-$\sigma$ detection thresholds for point sources are 29.4, 30.8, 30.2 and 31.1 mag for F275W, F475W, F814W and F160W respectively.

\begin{table*}
    \begin{center}
    \caption{\hst{} Observations of \eso{} (PI: Sun)}
    \label{tab:hstObs1}
    \vspace{-0.2cm}
    \begin{tabular}{cccccc}
        \hline
        Filter (mean $\lambda$ / FWHM) & Instrument & Mode & Dither $^\mathrm{a}$ & Date & Exp (sec) \\
        \hline
        F275W (2719 / 418 $\text{\AA}$) & WFC3/UVIS & ACCUM & 3 (2.4$''$) & 07/17/2011 & $3\times1010.0$  \\
        F475W (4802 / 1437 $\text{\AA}$) & ACS/WFC & ACCUM & 2 (3.01$''$) & 07/17/2011 & $2\times864.0$   \\
        F814W (8129 / 1856 $\text{\AA}$) & ACS/WFC & ACCUM & 2 (3.01$''$) & 07/17/2011 & $2\times339.0$ \\
        F160W (15436 / 2874 $\text{\AA}$) & WFC3/IR & MULTIACCUM & 2 (0.636$''$) & 07/17/2011 & $2\times399.2$  \\
        \hline
    \end{tabular}
    \end{center}
    \vspace{-0.2cm}
    Note: $^\mathrm{(a)}$ number of dither positions (and offset between each dither)
\end{table*}

The first step of the data reduction involves alignment of different frames in each filter using Tweakreg~\citep{2012drzp.book.....G}. 
For absolute astrometry, we used the Guide Star Catalog II~\citep[GSC2;][]{2008AJ....136..735L},
which aligns different frames within an error of $\sim 0.05''$.
After alignment, we ran AstroDrizzle to combine different frames for the same filter. 
The data were also corrected for charge transfer efficiency (CTE)
to ensure maximum accuracy in the photometry measurement. The pixel scale was fixed to 0.03$''$ (or 9.81 pc).

The combined images in the F275W, F475W and F814W bands still have many CRs left, 
especially at the ACS and WFC3 gaps, and edges. 
The Laplacian Edge Detection algorithm by \citet{2001PASP..113.1420V} has been used for further CR removal. The algorithm assumes that a real source such as a star or a galaxy has a smooth gradient light profile whereas a spurious source like CRs have discontinuous gradient. This algorithm usually works very well for ground optical data but requires extra caution and tests for the \hst{} data for its very sharp point spread function (PSF). We also ran SExtractor~\citep{1996A&AS..117..393B} with a low detection threshold for the sole purpose of identifying any spurious objects like CRs. Sources detected in only one {\em HST} band are considered candidate CRs. We visually inspected all these candidates to verify their nature as CRs and subsequently removed confirmed CRs from the image.

%% file: APEX_observ.tex
\section{{\em APEX} CO Observations}\label{sec:CO_observation}
The observations of \eso{} were carried out with the {\em APEX} 12\,m antenna in September 2011 with the program ID of 088.B-0934(A) and in September 2014 with the program ID of 094.B-0766(A). The observations were done at the frequency of the $^{12}$CO(2-1) ($\nu_{\rm rest} = 230.538$~GHz) using the APEX-1 receiver of the Swedish Heterodyne Facility Instrument (SHFI), and partially also at the $^{12}$CO(3-2) frequency ($\nu_{\rm rest} = 345.796$~GHz) using the APEX-2. The eXtended Fast Fourier Transform Spectrometer (XFFTS) backend was used with a total bandwidth of 2.5~GHz divided into 32768 channels. The corresponding velocity resolution is about 0.1~km\,s$^{-1}$. The XFFTS consisted of two units with a 1~GHz overlap region. It thus covered the entire IF bandwidth of the SHFI. At the observed CO(2-1) and CO(3-2) frequencies, the FWHM of the primary beam of the telescope is $\sim 26.7''$ and $\sim 17.8\arcsec$, respectively (following $\theta_{\rm MB}\approx 1.17 \lambda / D$), which for the adopted distance of the Norma Cluster corresponds to $\sim 8.7$~kpc and 5.8~kpc, respectively.

\begin{figure} 
    \includegraphics[width=0.475\textwidth]{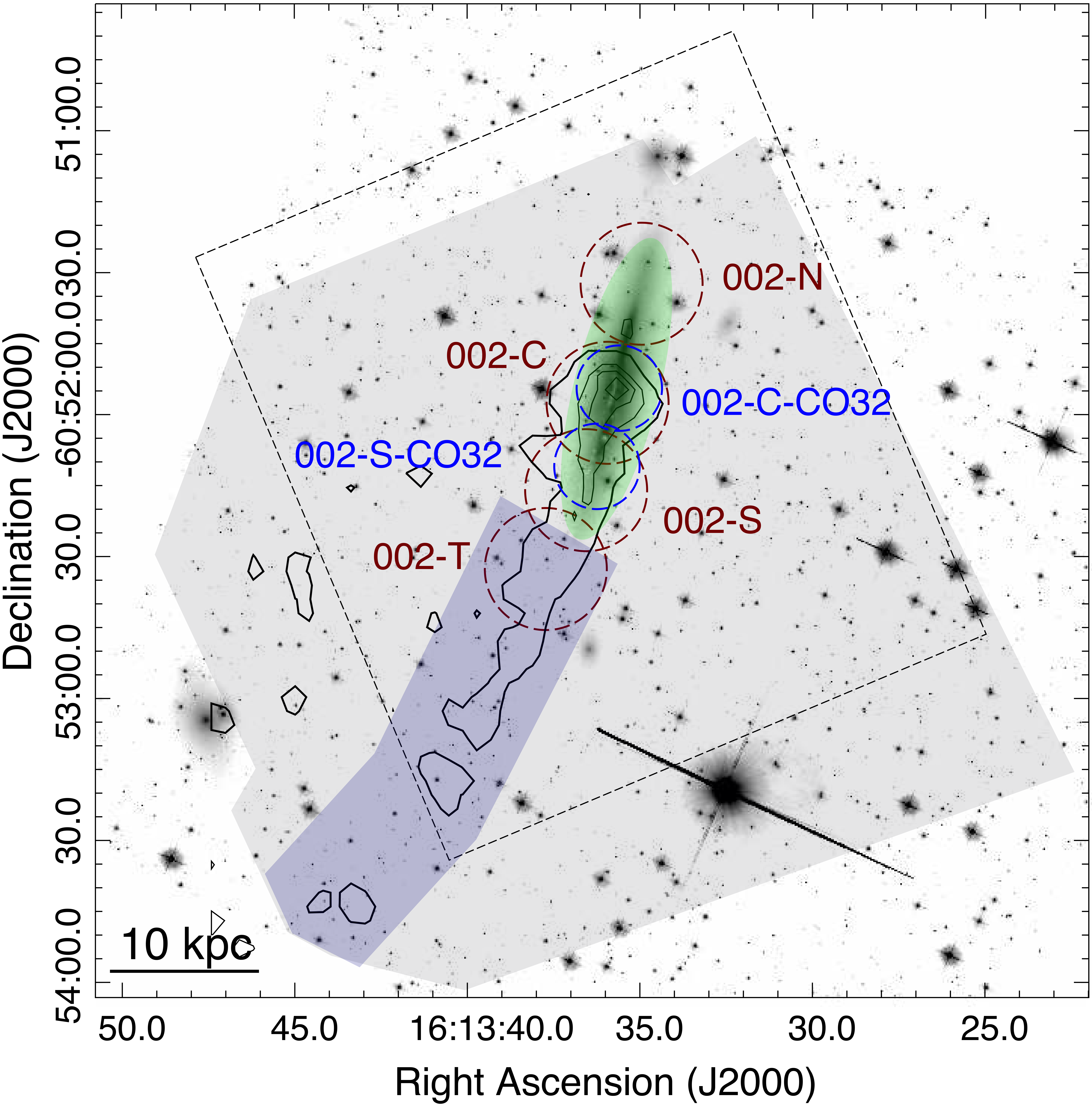}
    \vspace{-0.5cm}
    \caption{Regions of interest --- tail, galaxy, and control regions shown by a lightblue polygon, a green ellipse, and a grey polygon (excluding the first two regions), respectively, on the F475W image. The center and axial ratio of the ellipse is given in Section~\ref{sub:morph}. The tail region encloses the X-ray and H$\alpha$ tails shown in \citet{2013ApJ...777..122Z} and the X-ray contours are also shown. The choice of the control region is mainly decided by the FOV of the F275W data. The dashed line shows the FOV of the F160W data. The source populations in these three regions are studied in Section~\ref{sec:hst_srcs}.
    The red dashed circles show {\em APEX} pointings with $^{12}$CO(2-1) beams (${\rm FWHM}= 26''.7 = 8.7$~kpc) and the blue dashed circles show {\em APEX} pointings with $^{12}$CO(3-2) beams (${\rm FWHM}= 17''.8 = 5.8$~kpc). The CO spectra in these regions are shown in Fig.~\ref{fig:apex_galaxy}, Fig.~\ref{fig:apex_tail} and Fig.~\ref{fig:apex_co32} and discussed in section~\ref{sec:CO_emission}.
    }
    \label{fig:region}
\end{figure}

The observations were done in a symmetric Wobbler switching mode with the maximum throw of $300''$ in order to avoid with OFF positions the tail if oriented in azimuth. Three integration points were selected over the main body of the galaxy, one (``002-C'') aiming nearly at the center of \eso{} ($\sim 4.4''= 1.4$~kpc southwards off the center), another (``002-S'') covering the southern part of the disk, where H$\alpha$ and X-ray tail connects to the galaxy, and the last covering the northern disk part (see Fig.~\ref{fig:region}). The pointings 002-C and 002-S were selected to cover most of the brightest X-ray emission. The 002-N pointing is located symmetrically to the 002-S one (with respect to the optical galaxy center) in order to reveal possible asymmetries in the CO-emitting gas distribution. Another integration point (``002-T'') was aimed over the inner tail region, about $\sim 15$~kpc from its optical center, already outside of the disk. Two regions were also observed at the $^{12}$CO(3-2) frequency, over the optical center of the galaxy and an adjacent location to the south. Coordinates of the observed positions are given in Table~\ref{tab:apexObs}, together with the actual on-source observing times. The receiver was tuned to the $^{12}$CO(2-1) frequency shifted in each position to its respective optical radial velocity. Observing conditions were good with ${\rm PWV}\lesssim 1$~mm. The system temperatures were typically about 140~K.

\begin{table}
    \begin{center}
    \caption{APEX CO observations of \eso{} (PI: J\'{a}chym)}
    \label{tab:apexObs}
    \vspace{-0.2cm}
    \begin{tabular}{cccc}
        \hline
      Pointing (transition) & R.A. & Dec. & $T_{\rm ON}$ \\
        & (J2000) & (J2000) & (min)\\
        \hline
        002-C (2-1) & 16:13:35.83 & -60:51:58.7 & 26 \\
        002-S (2-1) & 16:13:36.47 & -60:52:16.8 & 43 \\
        002-N (2-1) & 16:13:34.89 & -60:51:32.3 & 89 \\
        002-T (2-1) & 16:13:37.70 & -60:52:33.6 & 77 \\
        \hline
        002-C-CO32 (3-2) & 16:13:35.62 & -60:51:54.6 & 42 \\
        002-S-CO32 (3-2) & 16:13:36.26 & -60:52:10.9 & 138\\
        \hline
    \end{tabular}
    \end{center}
    \vspace{-0.2cm}
\end{table}

The data were reduced according to the standard procedure using CLASS from the GILDAS software package developed at IRAM. Bad scans were flagged and emission line-free channels in the total width of about 1000~km\,s$^{-1}$ were used to subtract first-order baselines. The corrected antenna temperatures, $T^*_{\rm A}$, provided by the {\em APEX} calibration pipeline \citep{2010SPIE.7737E..1JD}, were converted to main-beam brightness temperature by $T_{\rm mb}=T^*_{\rm A} / \eta_{\rm mb}$, using a main beam efficiency of about $\eta_{\rm mb} =0.75$ at 230~GHz and 0.73 at 345~GHz. The rms noise levels typically of $1-2$~mK per 12.7~km~s$^{-1}$ channels were obtained in the main body pointings. Gaussian fits were used to measure peak $T_{\rm mb}$, width, and position of the detected CO lines. Flux densities can be obtained using the conversion factor $S_\nu/T_{mb}=39$~Jy\,beam$^{-1 }$/K for the APEX-1 receiver at 230~GHz, and 41~Jy\,beam$^{-1 }$/K for the APEX-2 receiver at 345~GHz.

%% file: photometry.tex
\section{\hst{} view of \eso{}}\label{sec:morph}
\subsection{Morphology \& Light Profiles} \label{sub:morph}

Fig.~\ref{fig:composite} shows the \hst{} composite and individual images of \eso{} in four bands, also with the X-ray image from {\em Chandra} shown. \eso{} is a large, edge-on spiral galaxy with a boxy bulge.
While the upstream side of the galaxy is nearly dust-free, there are many dust features downstream, which suggests that the RPS has nearly cleared the north half of the galaxy and the stripping is ongoing 
within 1 - 2 kpc of the nucleus and the southern part. We attempted to fit the F160W image of \eso{} (least affected by the dust extinction) with the two-dimensional image fitting algorithm --- GALFIT~\citep{2002AJ....124..266P}. Single and double S\'{e}rsic models are applied. The results are listed in Table~\ref{tab:galfit}. A double S\'{e}rsic model (one for the bulge and the other for the disk) can fit the galaxy image reasonably well. The resulting S\'{e}rsic index of each component is also reasonable with the trend found in large surveys like {\em GAMA}~\citep[e.g.,][]{2015MNRAS.447.2603L}.

\begin{figure}
    \vspace{0.1cm}
    \includegraphics[width=0.474\textwidth]{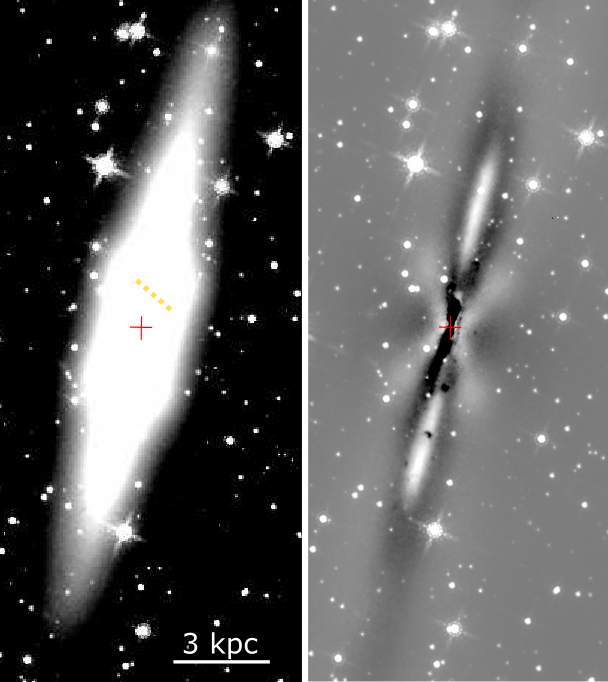}
    \vspace{-0.5cm}
    \caption{{\em Left}: the F160W image to show the boxy bulge. The yellow dashed line shows the approximate position of the stripping front, determined from the X-ray and H$\alpha$ data \citep{2013ApJ...777..122Z}. Note this projected stripping front is really the outer edge of the curved, 3D stripping front.
    {\em Right}: the residual image after removing the best-fit double S\'{e}rsic model with GALFIT (see Table \ref{tab:galfit} for the detail of the fit). The central bulge shows an X-shaped residual, typical for galaxies with a boxy bulge and a bar. The red cross in each panel shows the position of the nucleus.
    }
    \label{fig:eso_bulge}
\end{figure}

The galaxy has a boxy bulge (Fig. \ref{fig:eso_bulge}). After subtracting the double S\'{e}rsic model as shown in Table~\ref{tab:galfit}, the X/peanut-shaped bulge is clear, which implies the existence of a bar with a length of $\sim$ 5 kpc in the bulge.
The X-shaped structure can be the consequence of dynamical instabilities in the stellar bars creating thick vertical layers, often seen in massive galaxies~\citep[e.g.,][]{2017MNRAS.468.2058E, 2017A&A...598A..10L}.

\begin{figure*}
    \vspace{-0.5cm}
    \includegraphics[width=0.9\textwidth]{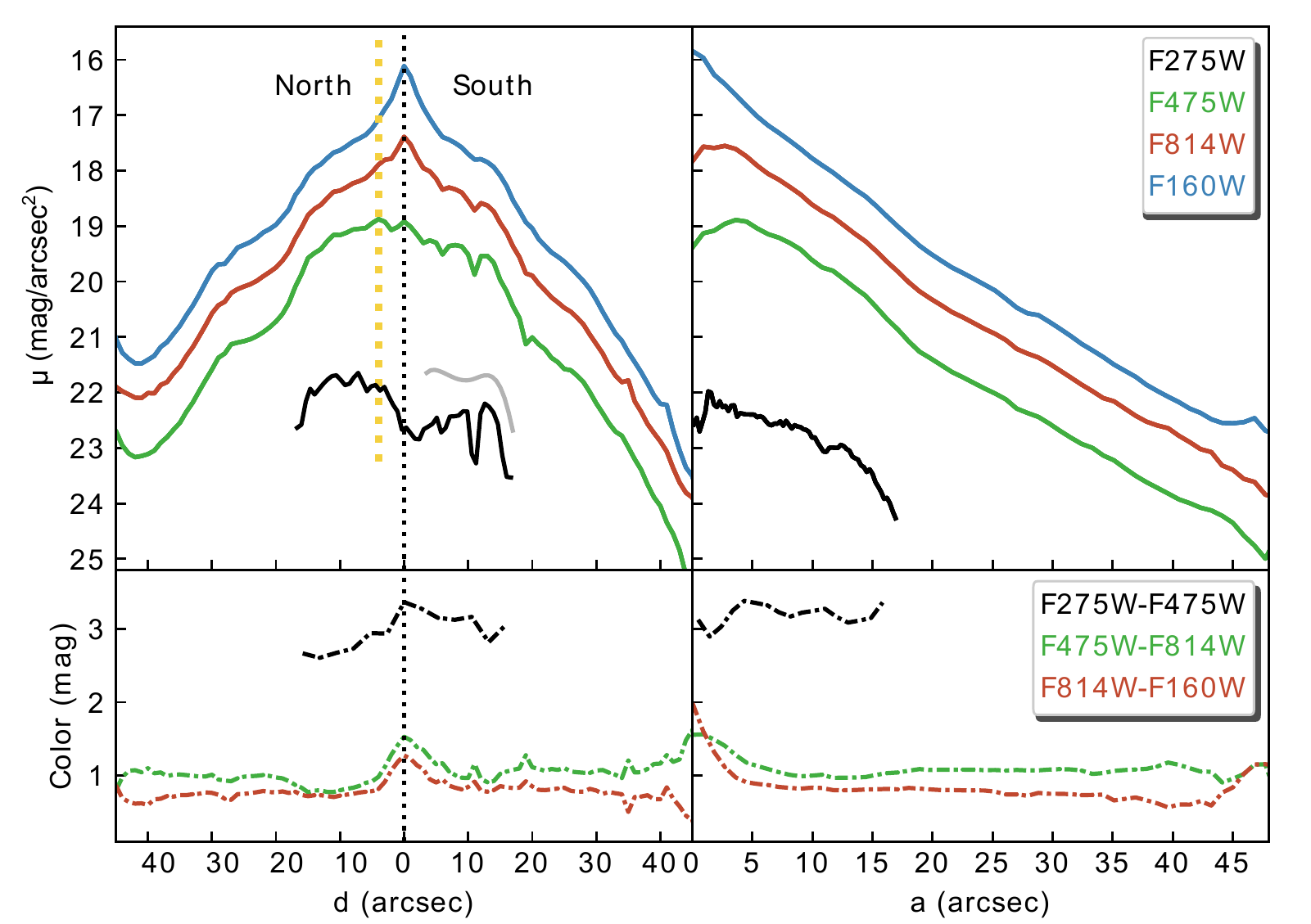}
    \vspace{-0.3cm}
    \caption{The surface brightness profiles measured along the major axis ({\em Upper Left}, with $d$ = 0 at the nucleus) and the elliptical annuli centered at the nucleus ({\em Upper Right}, $a$ as the semi-major axis), in all four bands. The color profiles are shown in the lower panel. The elliptical annuli have a PA of -13.1$^{\circ}$ and an axis ratio 0.23.
    One can observe the significant effect of dust extinction at short wavelengths, while the F160W (or $H$-band) profiles are smoother and more symmetric than the profiles of the other three bands. 
    As shown in Fig.~\ref{fig:composite}, the F275W light truncates at $\sim 16''$ from the nucleus. The yellow dashed line marks the position of the stripping front. The grey profile in the upper left panel is the downstream F275W profile after correction of the intrinsic extinction with the F475W and F814W data (see Section~\ref{sub:dust_features} for detail), which suggests that the galaxy would have been more symmetric in F275W without the intrinsic extinction (see Fig.~\ref{fig:composite}).
    }
    \label{fig:eso_profile}
\end{figure*}

To quantitatively examine the galaxy structure, we derive the surface brightness profiles in all four bands, along the major axis and in elliptical annuli centered at the nucleus (Fig. \ref{fig:eso_profile}). The galaxy center is set at the nuclear position presented in Section \ref{sub:nucleus}. The major axis has a position angle (PA) of -13.1$^{\circ}$ (measured from the north and running counterclockwise). The total F160W light of the galaxy is measured to be 11.39 mag (without correction on the intrinsic extinction) and the half-light semi-major axis is 3.30 $\pm$ 0.05 kpc at F160W. Both the total magnitude and the half-light size are similar to the best-fit results from the single-S\'{e}rsic fit from GALFIT. 

The light profiles along the major axis are measured in 25 boxes, each with a dimension of 2.0$''$ (width) $\times$ 2.5$''$ (length). The elliptical annuli have an axis ratio of 0.23.
Fig. \ref{fig:eso_profile} shows that the NIR light profiles in F814W and F160W are more symmetric and smoother than the blue light profiles in F275W and F475W, where the effect of dust extinction and SF is more pronounced. The strong dust lane around the nucleus affects all light profiles. The F275W light profile is the least smooth one. 
We also searched for tidal features around the galaxy, especially in the F814W and F160W bands. No such features are detected from the current data.

\begin{table}
    \begin{center}
    \caption{GALFIT fits on the F160W image of \eso{}}
    \label{tab:galfit}
    \vspace{-0.2cm}
    \begin{tabular}{cccc}
        \hline
        \hline
        Parameter  & Single & \multicolumn{2}{c}{Double}    \\ 
         &  ($\chi_{\nu}^{2}$=5.693) & \multicolumn{2}{c}{($\chi_{\nu}^{2}$=2.025)}\\
 \cline{3-4} 
    &      & Bulge   &  Disk   \\   
       
        \hline
        Total magnitude (mag)    & 11.53 & 12.49 & 12.12  \\
        $r_e$ (kpc)                 & 4.09 &  2.05 & 5.17 \\
        S\'{e}rsic index             & 1.41  &  2.43 & 0.74\\
       Axis ratio                    & 0.23  & 0.66  & 0.13 \\
        PA ($\deg{}$) & -13.1 & -13.9 & -13.3 \\
        \hline
        \hline
    \end{tabular}
    \end{center}
    \vspace{-0.2cm}
    Note: The axis ratio is the ratio between the minor axis and the major axis. The position angle is measured relative to the north and counter-clockwise.
\end{table}

\begin{figure}
    \vspace{0.1cm}
    \includegraphics[width=0.49\textwidth]{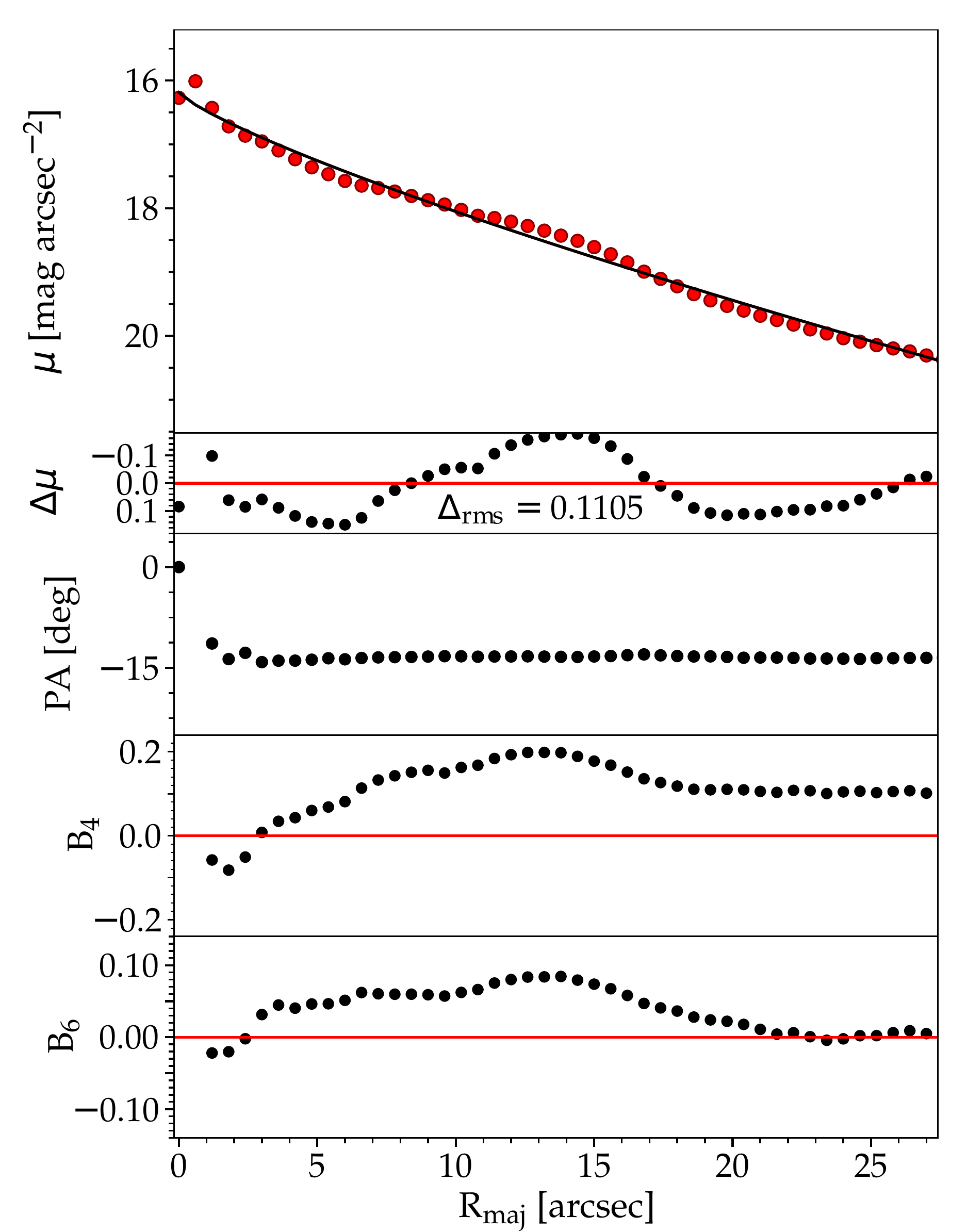}
    \vspace{-0.5cm}
    \caption{
    The results from PROFILER on the F160W image of \eso{}.
    The top panel shows the brightness profile (red circle) with the semi-major axis~($R_{\text{maj}}$). The profile is modeled with a single S\'{e}rsic component~(black solid line) which yields a S\'{e}rsic index of 1.24 and a half-light radius of 4.36 kpc, in good agreement with the GALFIT result~(see Table~\ref{tab:galfit}). Directly underneath is the residual profile followed by the PA profile,
    the fourth cosine harmonic amplitude $B_{4}$ and the sixth cosine harmonic amplitude $B_{6}$ profiles. The $B_{4}$ profile indicates the presence of a boxy bulge at $R_{\text{maj}} < 3''$ while the $B_{6}$ profile confirms the existence of an X/peanut-shaped bulge. 
    }
    \label{fig:eso_profiler}
\end{figure}

We further applied PROFILER~\citep{2016PASA...33...62C} to decompose the 1D light profile to verify the non-axisymmetric components like an X/peanut-shaped bulge in \eso{}. 
The input isophote table was generated by the IRAF task ISOFIT~\citep{2016MNRAS.459.1276C} which is appropriate for the edge-on disc galaxies. We fit the F160W image with a single S\'{e}rsic component while keeping the ellipticity  
and PA free. The best fitted values for the S\'{e}rsic index and the half-light radius are 1.24 and 4.36 kpc, respectively, in good agreement with the GALFIT results. Fig.~\ref{fig:eso_profiler} shows the full light profile along with the residual, PA and two harmonic namely the fourth cosine harmonic amplitude~($B_{4}$) and the sixth cosine harmonic amplitude~($B_{6}$). The  $B_{4}$ profile shows that the galaxy appears boxy within $R_{\text{maj}} \sim 3''$ and becomes disky at large radii.
The $B_{6}$ profile shows the significance of the X/peanut-shaped component at $R_{\text{maj}} \sim 5''$ - 16$''$.

\subsection{Dust Features}\label{sub:dust_features}
\begin{figure*}
    \includegraphics[width=1.0\textwidth]{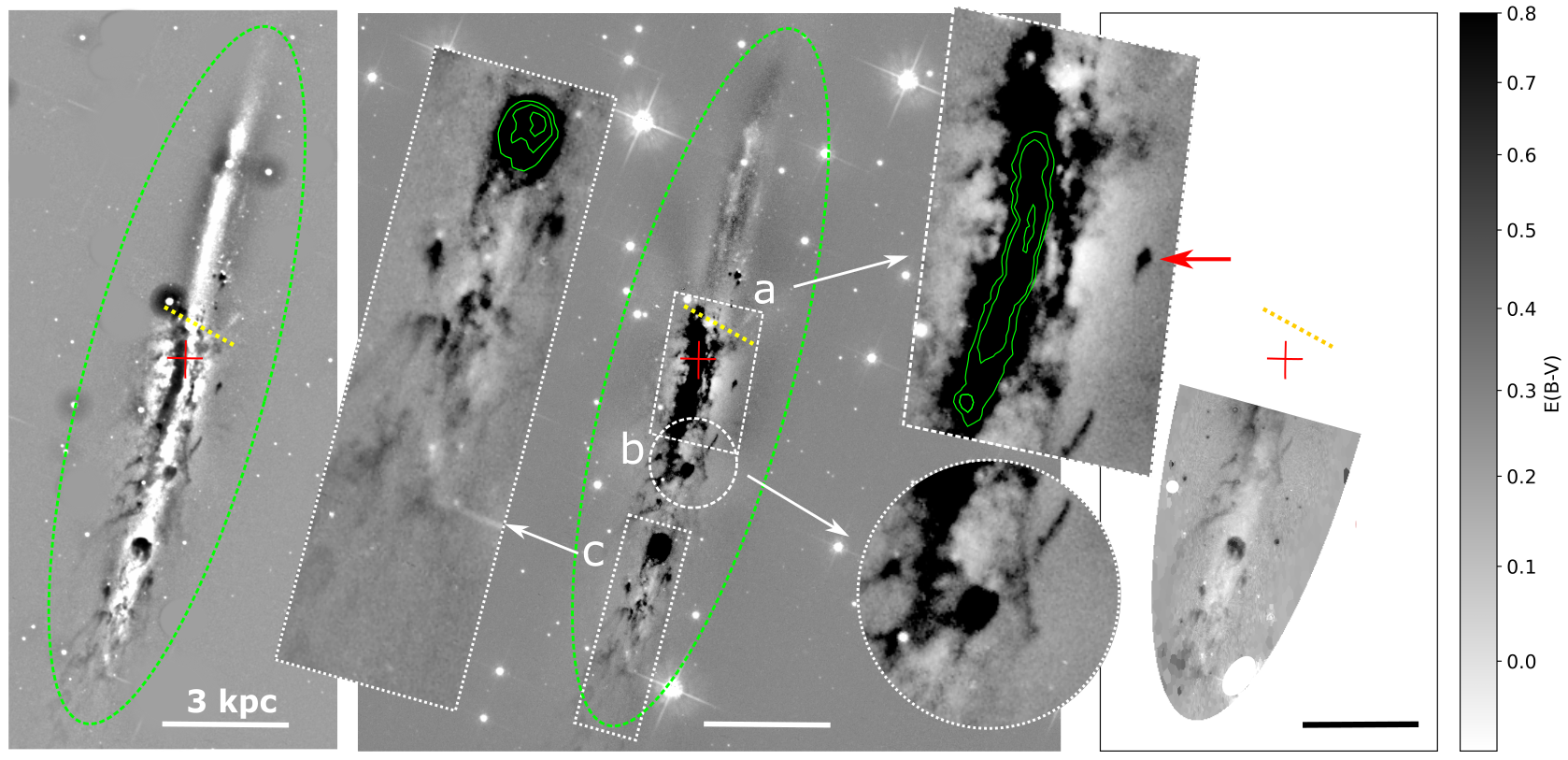}
    \vspace{-0.6cm}
    \caption{{\em Left}: The unsharp masked F475W image of \eso{}, after subtracting the original image smoothed with 1 pixel (or 0.03$''$) from the same image smoothed with 20 pixels. Prominent dust lanes and filaments are visible which extend downstream from the nucleus. The red cross shows the position of the nucleus (the same for the other two panels) and the green ellipse shows the region for the E(B-V) analysis (the same for the middle panel). The yellow dashed line is the same stripping front as shown in Fig.~\ref{fig:eso_bulge} (also the same for the other panels).
    {\em Middle}: the residual F475W image, after subtracting the GALFIT model image obtained with four S\'{e}rsic components with the Fourier mode (see Section \ref{sub:dust_features}). Similar dust features as shown on the left are also visible in this residual image and highlighted in zoomed images from a to c (the large and smooth dark features upstream are from over-subtraction of the model, instead of real dust features).
    Inset a) shows the main dust lane around the nucleus, with the green contours show the levels of deficits. Inset b) shows some dust filaments trailing the SE of the nuclear region.
    Inset c) shows a large dust cloud (with a radius of $\sim$ 0.22 kpc) in the middle of downstream (green contours also show the levels of deficits), as well as many dust clouds and filaments further downstream that can be traced up to $\sim$ 13.5 kpc from the nucleus. The isolated dust cloud pointed by the red arrow in the inset a and the X-shaped feature in the inset b are likely clouds in fallback (see text for discussion).
    {\em Right}: The E(B-V) map derived from the F475W and F814W images, assuming that the upstream region is dust free and an extinction law from~\citet{1999PASP..111...63F}.
    }
    \label{fig:galfit_unsharp}
    \end{figure*}

To better show the dust features in the galaxy, we applied unsharp masking on the F475W image, by subtracting the original image smoothed with 1 pixel from the same image smoothed with 20 pixels (Fig.~\ref{fig:galfit_unsharp}). 
It shows many dust features around the nucleus and downstream, extending to at least 13.5 kpc to the SE. A prominent dust lane is observed around the nucleus and there are many dust filaments and clouds downstream. One can also see the upstream end of the nuclear dust lane ($\sim$ 1.5 kpc from the nucleus) is pushed to the east side, presumably by the ram pressure. The downstream part of the nuclear dust is also pushed to the east side (but to a less extent compared with the upstream part). Assuming a galactic rotation speed of 100 km/s at 1.5 kpc radius from the nucleus, the downstream and upstream dust lanes and the associated gas clouds would switch positions in $\sim$ 46 Myr. Is the difference on the curvature of the dust lanes upstream and downstream caused by the cloud fallback once moved to the downstream that is less affected by ram pressure? Future spectroscopic data with e.g., {\em MUSE} or {\em ALMA} can examine the ISM kinematic under the ram pressure impact.

We also used GALFIT again to produce a smooth model of the galaxy, this time on the F475W image. The final GALFIT model has four S\'{e}rsic components, two in the disk and two in the bulge, along with the Fourier modes (mode= 4, 6 and 8)~\citep{2010AJ....139.2097P}.
The residual image is shown in the middle panel of Fig.~\ref{fig:galfit_unsharp}. It should be noted that the dark features upstream of the galaxy are caused by the over-subtraction of the model and not caused by dust. Its smooth and extended distribution is clearly different from those of local dust clouds.
We also highlight dust features in zoom-in insets. Inset a) shows the strong dust lane in the bulge, with the nucleus buried in the position with the strongest dust extinction. Inset b) goes further downstream and encloses an interesting 'X'-shaped feature with a length of $\sim$ 0.8 kpc.
Inset c) shows numerous dust threads, arcs and filaments further downstream, to at least 13.5 kpc from the nucleus.
These dust threads have a typical length of $\sim$ 0.8 kpc and width of $\sim$ 0.1 kpc, which can be compared with dust filaments in the leading edge of NGC 4921~\citep{2015AJ....150...59K} with 0.5 - 1 kpc in length and 0.1 - 0.2 kpc in width.
The dust filaments in the downstream are roughly aligned with the RPS direction, determined from the X-ray/H$\alpha$ tails. This reflects the effect of RPS on dust ablation from dense ISM clouds. 
Most dust filaments around the nucleus and downstream have angles of $\sim$ 20$^\circ$ from the major axis of the galaxy, which is taken as the 2D wind angle.
The orientation of these dust filaments may also be affected by magnetic field and disk rotation in \eso{}.
The significance of these dust features also suggests that they mostly lie in the front side of the galaxy closer to us.
The existence of long, coherent dust filaments suggests that magnetic field plays an important role here.
If a similar dust-to-gas ratio as in the Milky Way is assumed \citep{2009MNRAS.400.2050G}, the \ion{H}{I} + H$_2$ column density of the strongest dust filament is $\sim$ 13.7 M$_{\odot}$ pc$^{-2}$ at the downstream region.

Interestingly, several dust clouds far from the major axis (e.g. one marked by a red arrow and an X-shaped feature in Fig~\ref{fig:galfit_unsharp}) are not near the disk mid-plane but are extraplanar on the ``wrong side'' (upstream side), as the ICM wind has a west component. With an inclination angle of $\sim$ 83$^\circ$ and a PA of $\sim$ -13.1$^\circ$, the cloud marked by the red arrow in Fig.~\ref{fig:galfit_unsharp} is at $\sim$ 0.8 kpc from the disk mid-plane. These clouds likely present evidence of gas fallback as predicted in RPS simulations \citep[e.g.,][]{2012MNRAS.422.1609T}.
Clouds are initially accelerated upwards from the disk. However, without achieving escape velocity, at least some will ultimately fall back, ending up momentarily on the upstream side of the disk. One expects dense clouds to be more prone to fallback since they are accelerated less with the same ram pressure.

To quantify the dust extinction in the galaxy, we derive the E(B-V) map of the galaxy, with the F475W and F814W data. A key assumption in this analysis is that the F475W - F814W color of the galaxy is symmetric around the nucleus.
Since the upstream appears dust-free (e.g., beyond 5$''$ from the nucleus in Fig. \ref{fig:eso_profile}), we can use the upstream color as the color in the mirror position of the downstream without intrinsic extinction. We employ the Voronoi binning method~\citep{2003MNRAS.342..345C} to adaptively bin the F475W image with a minimum S/N of 50 and also bin the F814W image with the same choice of bins. The new binned images were then converted to an E(B-V) map using the equation

\begin{equation}\label{eq:EBV}
    E(B-V) = - \frac{(F475W - F814W) - (F475W - F814W)_\circ}{k_{\text{F475W}} - k_{\text{F814W}}}
\end{equation}

F475W - F814W is the measured color downstream, while (F475W - F814W)$_{\circ}$ is the measured color in the corresponding upstream position, taken as the color without the intrinsic extinction. $A_\lambda = k_\lambda E(B-V)$ is the extinction law.
Since \eso{} is a $L_*$ spiral galaxy like our Milky Way, we use the extinction law of~\citet{1999PASP..111...63F} with R$_V$ = 3.1. The results are shown in Fig. \ref{fig:galfit_unsharp}, with indeed enhanced extinction at the positions of dust clouds.

With this correction, it appears that the F275W light distribution becomes more symmetric between downstream and upstream regions. Thus, the SF is not strongly enhanced in the downstream regions, although some evidence of localized SF triggered by RPS is observed downstream as shown in the next subsection. 

\subsection{Triggered SF in the galaxy}\label{sub:triggerSF}

\begin{figure*}
\vspace{-0.1cm}
    \includegraphics[width=0.99\textwidth]{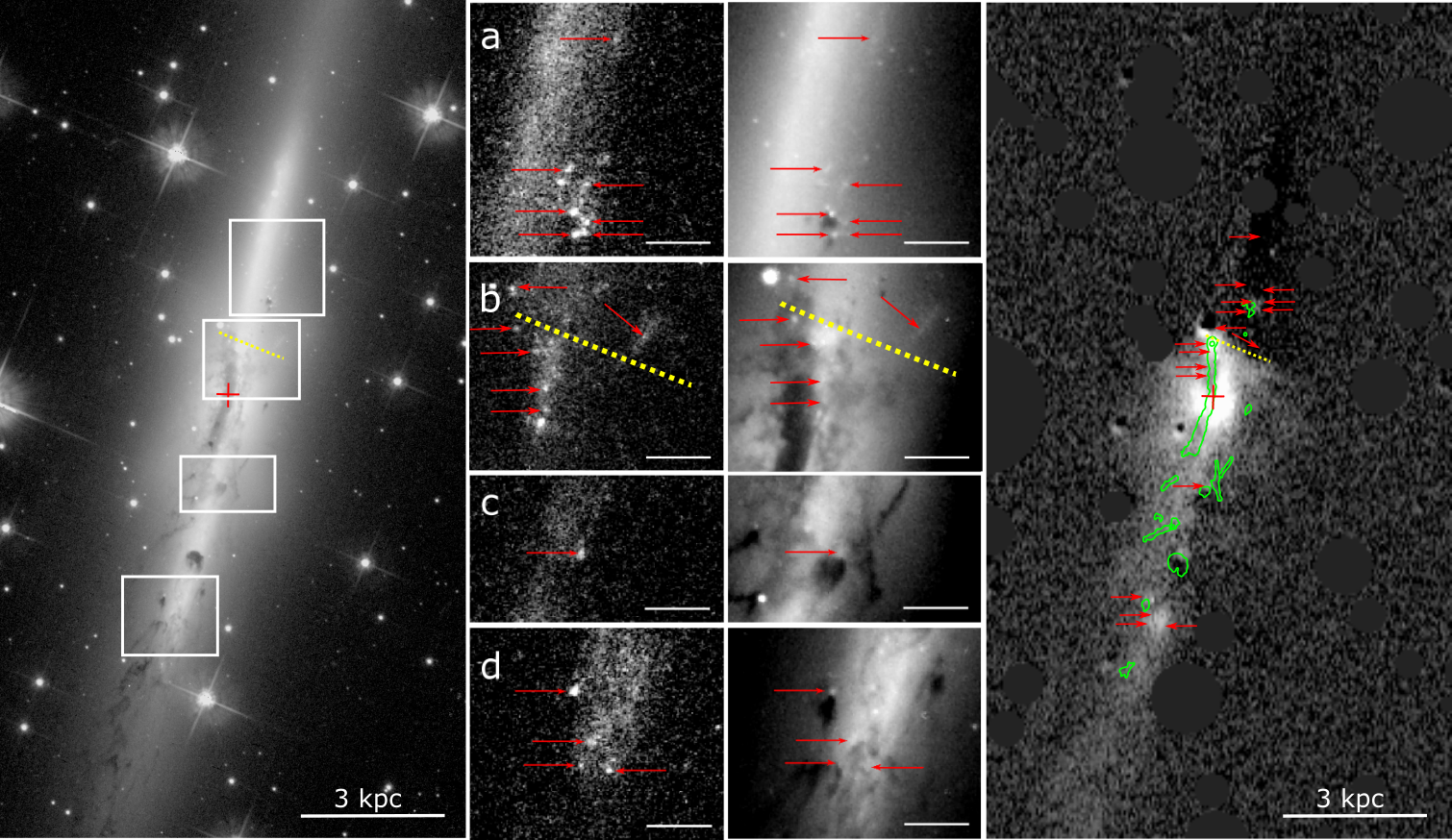}
    \vspace{-0.2cm}
    \caption{
    UV-bright young star clusters in the galaxy.
    {\em Left:} The F475W image of \eso{} with four white boxes to zoom in around young star clusters revealed in the F275W band. 
    The stripping front is shown by the yellow dashed line (the same as in Fig.~\ref{fig:eso_bulge}). Most F275W-bright compact sources are found around the stripping front.
    {\em Middle:} The four zoom-in regions (a, b, c and d) to show the young star clusters in the disk (highlighted by red arrows) in the F275W ({\em left}) and F475W ({\em right}) images. These young star clusters are bright in F275W but faint in F475W and F814W. Some are located immediately upstream of dusty clouds, which implies that the SF there was triggered by the compression from ram pressure. The white scalebar in each small panel is 0.5 kpc.
    {\em Right:} The net H$\alpha$ image of \eso{} from {\em SOAR} \citep{2010ApJ...708..946S}, with regions around bright stars masked and also overlaid with contours of significant dust filaments in green. The same UV-bright sources marked in the middle panels are also marked here, as well as the stripping front and the nucleus. The upstream of the galaxy is also free of the H$\alpha$ emission. There is a faint H$\alpha$ source at the position of the young star cluster association just upstream of the stripping front as shown in the small panel a.
    }
    \label{fig:young_star_region}        
\end{figure*}

While the galaxy is much fainter in F275W than in F475W (Fig.~\ref{fig:composite} and Fig.~\ref{fig:eso_profile}), there are some compact knots and clumps revealed in F275W with blue colors as shown in Fig.~\ref{fig:young_star_region}. Several blue sources bright in F275W also have enhancement in the {\em SOAR} H$\alpha$ image, likely \ion{H}{II} regions. These are most likely young star clusters formed recently. Interestingly, all of these blue sources have dust clouds nearby and some of them are immediately upstream of dust clouds (Fig.~\ref{fig:young_star_region}). This association suggests that the star formation of at least some of these young star clusters (if not all) is triggered by the compression of ram pressure.
Not all dust clouds have associated blue star clusters, which may reflect their different evolution stages.

On the other hand, these young star clusters may just shine in F275W because ram pressure has cleared dust around them, without actually triggering their formation. To test this scenario, we can move the F275W sources atop the cloud into the cloud to see whether they can still be detected. In principle, when a uniform dust cloud is put in front of a young star cluster and its surroundings, the brightness of the young star cluster and the surrounding r.m.s. will be reduced by the same factor so the significance of the young star cluster remains the same in this ideal situation. We further examined two sources, one in Fig.~\ref{fig:young_star_region} - c and another as the brightest one in d. These two sources are detected at 3.8 $\sigma$ and 6.0 $\sigma$ in F275W, respectively. With the $E(B-V)$ map in the galaxy derived in section~\ref{sub:dust_features}, we can examine the expected significance of these two sources if moving them to the center of their associated clouds. After adjusting the difference on extinction, these two sources would have been detected at 3.0 $\sigma$ and 4.3 $\sigma$,respectively, in F275W at the center of their associated clouds.
As no UV-bright star clusters are detected around the center of dust clouds (even those with low E(B-V) values) and few of them are detected in regions completely free of dust, the observed correlation between some young star clusters and the dust clouds is better explained by the triggered SF from ram pressure. Similar RPS-triggered SF was also observed in e.g., NGC~4921 \citep{2014ApJ...780..119K}, NGC~4402 and NGC~4522 \citep{2014AJ....147...63A,2016AJ....152...32A}, UGC~6697 \citep{2017A&A...606A..83C}, and IC~3476~\citep{2021A&A...646A.139B}.

Most of these young clusters are downstream of the stripping front, as the upstream or north side of the galaxy has been cleared by ram pressure. The stripping front is defined by the H$\alpha$ and X-ray emission front as the region upstream of the stripping front is free of H$\alpha$ and X-ray emission, also with few dust features.
\footnote{This is really a projected stripping front as the dimension of the front along the line of sight is substantial. Moreover, the front can be porous to allow ram pressure, at a direction somewhat tilted from the disk plane, to reach downstream regions.}
However, there are still some young star clusters upstream of the front, including several around the only remaining dust cloud and a blue stream just beyond the stripping front (the first two zoom-in boxes in Fig.~\ref{fig:young_star_region}).
The large number of young star clusters immediately upstream of the stripping front also supports the scenario of SF triggered by ram pressure.

\begin{figure*}
\vspace{-0.3cm}
    \includegraphics[width=0.97\textwidth]{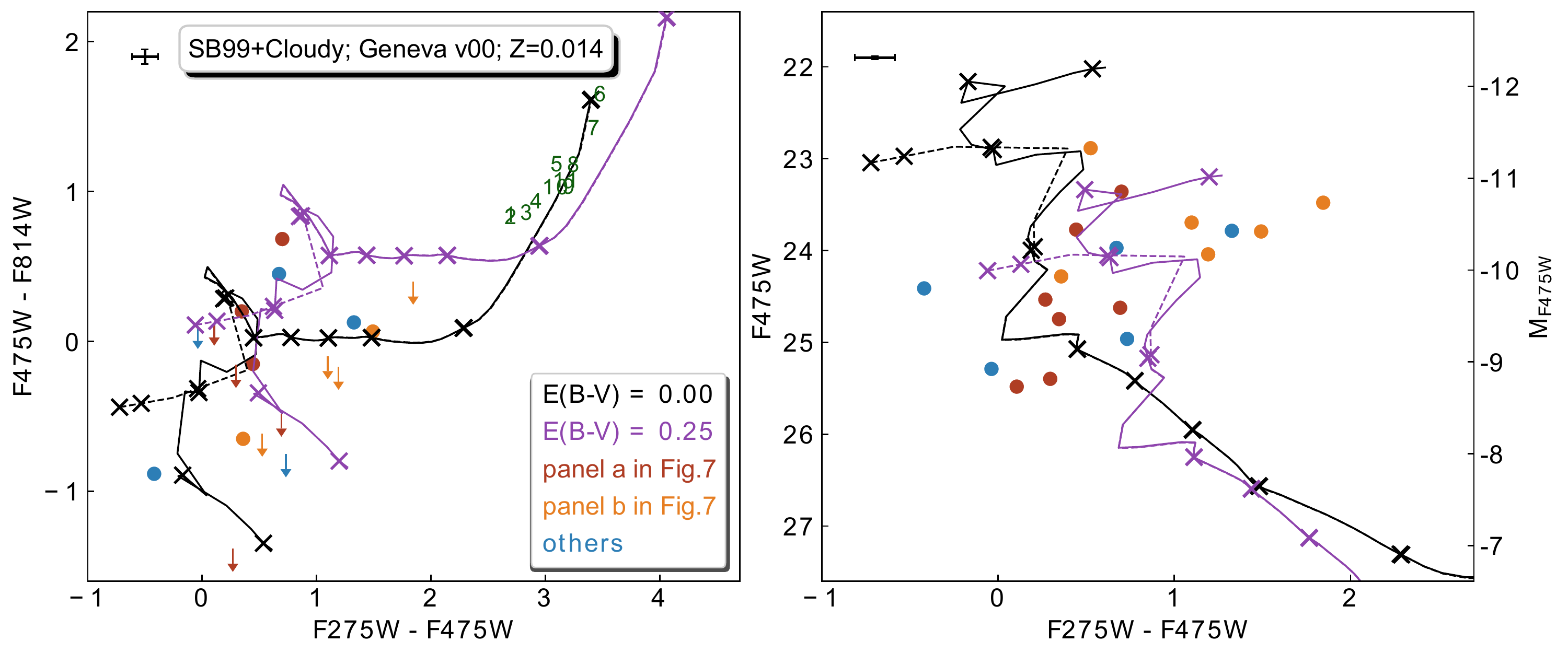}
    \vspace{-0.3cm}
    \caption{
    {\em Left}: The color-color diagram for blue sources in the disk identified in Fig.~\ref{fig:young_star_region}, with the typical errors of colors are shown on the upper left corner. 
    Sources in the region b are generally redder than those in other regions, which may be at least partially due to the enhanced intrinsic extinction around the nuclear region. Sources in the region a and other regions (c and d) have similar colors.
    A SB99 model with the Geneva v00 track is shown by the dashed line for two values of the intrinsic extinction, E(B-V)=0 and 0.25.
    We also ran Cloudy to include nebular emission that is important at ages of $<$ 10 Myr, and added the nebular emission on the SB99 tracks as shown by the solid line. The markers on the track indicate ages of 1, 3, 5, 10, 30, 50, 100, 200, 500 and 1000 Myr, starting from the blue end (dashed and solid lines are the same at ages $> $ 10 Myr).
    The galactic disk shown in the F275W image (Fig.~\ref{fig:composite}) is divided into 11 boxy regions (a length of 3.0$''$ $\times$ a width of 2.6$''$), starting from \#1 in the north (upstream) to \#6 at the nucleus and \#11 in the south (downstream). The color in each box is shown by green number. It is clear that these blue sources selected in F275W are bluer than the diffuse galactic disk also defined in F275W. These tracks suggest that the blue sources have ages of several Myr to $\sim$ 200 Myr, while the diffuse galactic disk has an age of 500 - 1000 Myr if fit with a single-burst model. 
    {\em Right}: The color-magnitude diagram for the same sources with the absolute magnitude also shown on the right side, with the typical errors of colors are shown on the upper left corner. The same SB99 tracks are also shown (for a total mass of $10^4$ M$_\odot$) for two E(B-V) values and with/without the nebular emission from Cloudy. The markers on tracks are also the same but only run to 500 Myr. From these tracks, the mass of these young star clusters is 10$^{3} - 10^{4}$ M$_{\odot}$ if younger than $\sim$ 100 Myr.
    }
    \label{fig:young_star_color_mag}        
\end{figure*}

Source detection and aperture photometry were performed in F275W, F475W and F814W with SExtractor, with an aperture radius of 0.5$''$.  The ``dual image mode'' was used, with the F275W data as the detection. The AB magnitude system is used in this work.
Diagrams of the F275W - F475W color vs. the F475W - F814W color, as well as the F275W - F475W vs. the F475W magnitude for these sources, are shown in Fig.~\ref{fig:young_star_color_mag}.
As a comparison, we also measured the colors of the diffuse galactic disk detected in F275W, by defining 11 boxes along the major axis of the disk, each with a size of 3.0 kpc $\times$ 2.6 kpc excluding all unresolved sources. The nomenclature of these boxes has \#1 as the northernmost one (most upstream), \#6 at the nucleus and \#11 as the southernmost one. Their colors are also shown in Fig.~\ref{fig:young_star_color_mag}. It is clear that the compact sources that are bright in F275W are bluer than the diffuse galactic disk. On the other hand, the upstream portion of the diffuse disk is bluer than the downstream portion, with the nuclear region the reddest. This is caused by the intrinsic extinction around the nucleus and downstream, as the upstream is nearly clear of gas and dust.
 
We can compare colors of these sources with simple stellar population~(SSP) models.
Here we use the Starburst99 (SB99 hereafter) model \citep{1999ApJS..123....3L} with the Kroupa initial mass function (IMF)~\citep{2001MNRAS.322..231K} and Geneva~2012 zero rotation tracks~\citep{2012ApJ...751...67L, 2014ApJS..212...14L} to derive the track of star clusters, with a metallicity of 0.014 (or 0.7 solar, appropriate for \eso{}).
Instantaneous SF is assumed.
Pysynphot6 \citep{2013ascl.soft03023S, 2014MNRAS.442..327B} was used.
Fig.~\ref{fig:young_star_color_mag} shows two tracks with and without intrinsic extinction overlaid on the sources.

As some young star clusters in the disk may be quite young, their nebular emission can be significant. To account for the contribution of the nebular lines into the broad \hst{} bands, we employed the development version of the photoionization code Cloudy, last reviewed by \citet{2017RMxAA..53..385F}, to add nebular emission to the stellar component of the radiation field reported by SB99. Cloudy does a full ab initio simulation of the emitting plasma, and solves self-consistently for the thermal and ionization balance of a cloud, while transferring the radiation through the cloud to predict its emergent spectrum.
We assumed a nebula of density 100~cm$^{-3}$, and metallicity of
0.7 solar, surrounding the stellar source and extending out to 1~kpc from it. For the inner radius of the cloud, we experimented with two values (1~pc and
10~pc), but found that the predicted colors do not depend on that choice.
We also imposed a lower limit of 1\% on the electron fraction to let the
calculation extend beyond the H{\sc ii} region, into the photo-dissociation
region. The Cloudy modification is only important for star clusters younger than 10 Myr (Fig.\ref{fig:young_star_color_mag}) but does help to explain the blue color of F475W - F814W for some sources.

As shown in Fig.\ref{fig:young_star_color_mag}, the blue sources highlighted in Fig.\ref{fig:young_star_region} have ages of several Myr to about 200 Myr, depending on the amount of intrinsic extinction. There is also no significant difference between sources upstream and downstream of the stripping front, which may suggest different timescales of SF after the initial ram pressure compression for different clouds, on top of different intrinsic extinction. 
With an age range of 3 - 100 Myr, the estimated total mass of these young star clusters are $\sim$ 4.3$\times10^5$ M$_{\odot}$ and the corresponding SFR is $\sim 1.4\times10^{-3}$ $\text{M}_{\odot}$/yr which is significantly lower than the total SFR of the galaxy. When the intrinsic extinction with E(B-V)=0.25 mag (still assuming the extinction law from \citealt{1999PASP..111...63F}) is included, both the total mass and the SFR would increase by $\sim$ 3 times.

In contrast, the galactic disk holds an older population of stars~(500 - 1000 Myr) that evolved before the galaxy plunged into the dense ICM.
It is also noted that no compact F275W sources are detected beyond $\sim$ 1.2 kpc upstream of the stripping front, which constrains the speed of the stripping front. If the age of the star cluster association in region a of Fig.~\ref{fig:young_star_region} is 10 - 30 Myr, the stripping front would have moved with a speed of 120 - 40 km/s, if SF happens immediately after ram pressure compression. Since SF is almost certainly delayed after the initial compression, the actual speed of the front will be smaller.

\subsection{Nucleus}\label{sub:nucleus}

\eso{}'s nucleus hosts an obscured, Seyfert2-like AGN from the X-ray data~\citep{2010ApJ...708..946S,2013ApJ...777..122Z}. 
The nuclear position is determined from the {\em Chandra} data, at (16:13:35.763, -60:51:54.74) with an uncertainty of $\sim$ 0.5$''$.
The \hst{} data reveal that the nuclear region is indeed heavily obscured by dust, without a nuclear point-like source in the optical and NIR (Fig.~\ref{fig:composite}, Fig.~\ref{fig:galfit_unsharp} and Fig.~\ref{fig:young_star_region}). The average E(B-V) values around the nucleus is $\sim$ 0.5.

%% file: HST_src_pop.tex
\section{\hst{} source population in the tail}\label{sec:hst_srcs}

We also want to examine the source population in the X-ray/H$\alpha$ tail and search for young star clusters as found in many ``jellyfish'' galaxies. As shown in Fig.~\ref{fig:region}, we defined three regions of interest --- tail, control and galaxy regions. The tail region encloses the X-ray and H$\alpha$ tails as shown in \citet{2013ApJ...777..122Z}. 
The choice of the control region is mainly determined by the common FOV between the F275W and the F475W/F814W data. The galaxy region encloses sources in the galactic disk. The region area, after excluding bright stars, is 0.747, 4.97 and 0.281 arcmin$^{2}$ for the tail, control and galaxy region, respectively.

\begin{figure*}
\vspace{-0.3cm}
    \includegraphics[width=0.65\textwidth]{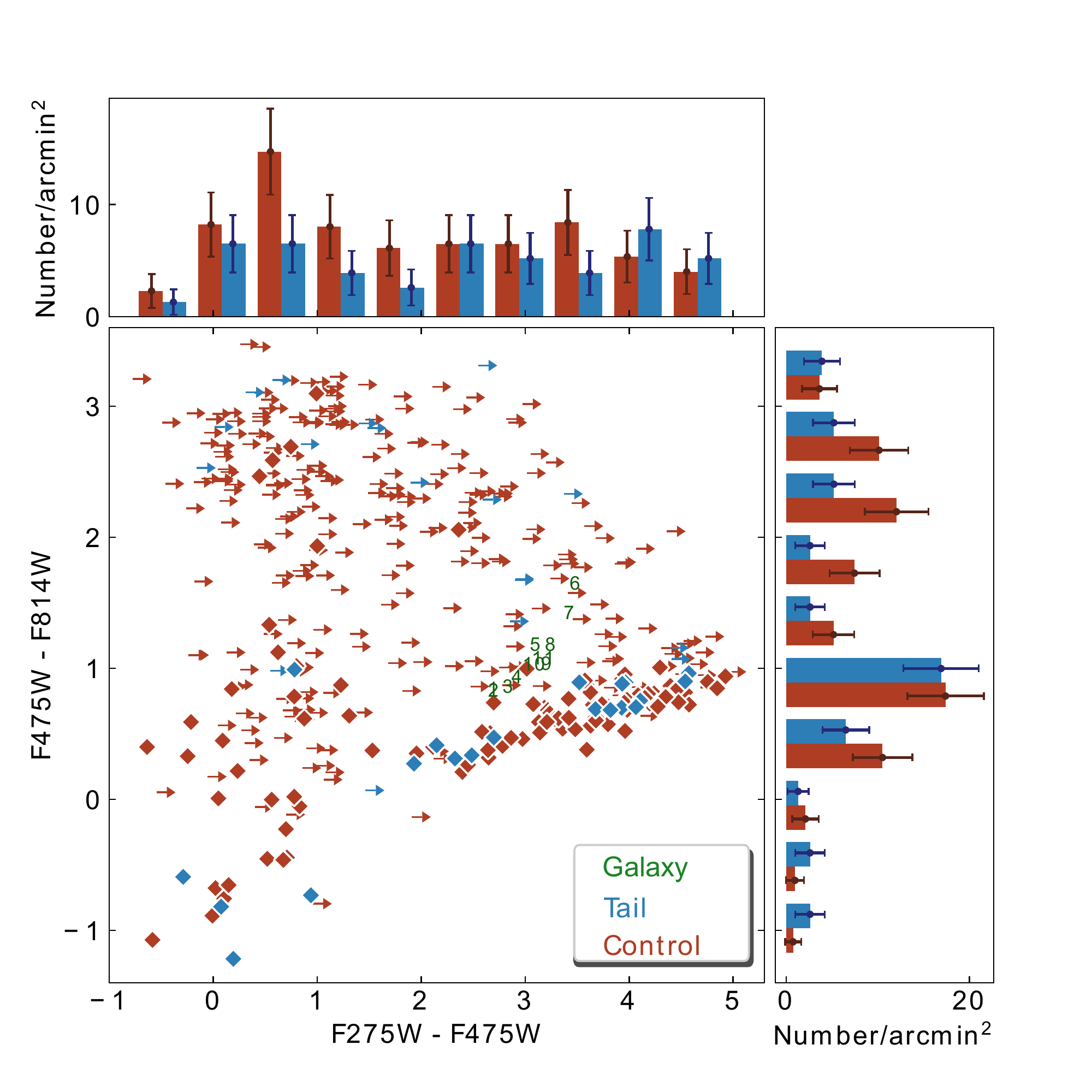}
    \vspace{-0.6cm}
    \caption{The color-color diagram for sources detected in F275W, F475W and F814W (diamonds) and sources only detected in F475W and F814W (arrows), in tail (blue) and control (red) regions. The initial detection was made in F475W. Most sources are undetected in F275W so F275W $-$ F475W colors are mostly limits. The tail/control regions are defined in Fig.~\ref{fig:region}. The source selection is discussed in Section~\ref{sec:hst_srcs}. 
    Colors of the galactic disk are again shown in green numbers as in Fig.\ref{fig:young_star_color_mag}.
    The median F475W $-$ F814W errors are 0.02 mag and 0.06 mag for the tail and control regions, respectively. The histograms are normalized to the region area (0.747 arcmin$^{2}$ for the tail region and 4.97 arcmin$^{2}$ for the control region). Error bars of the source counts are also plotted. As the distributions of both colors in the tail region are consistent with those in the control region, there is no evidence of young star clusters in the tail region.
    }
    \label{fig:hist_scatter}        
\end{figure*}

\begin{figure*}
\vspace{-0.3cm}
    \includegraphics[width=0.65\textwidth]{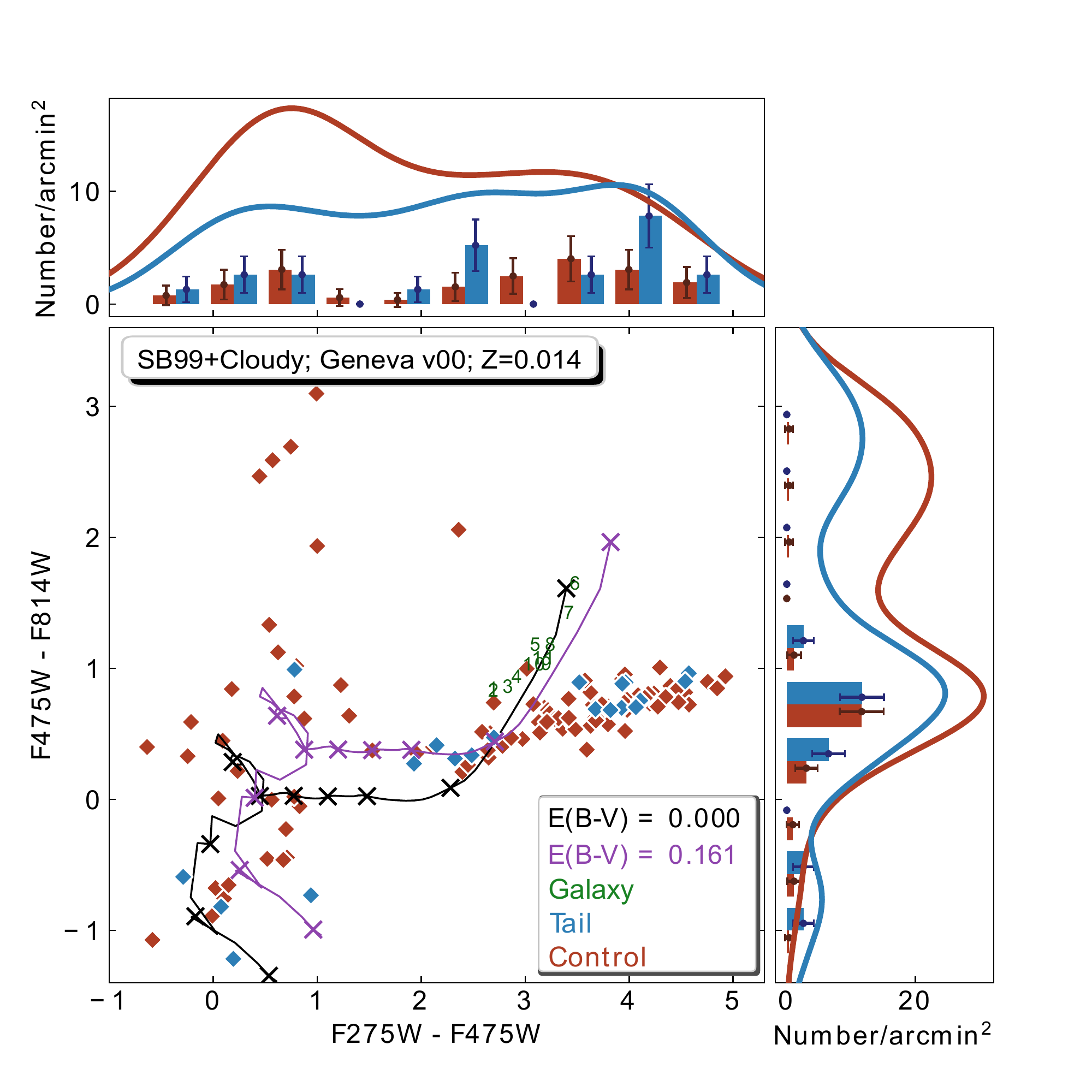}
    \vspace{-0.6cm}
    \caption{Similar to Fig.~\ref{fig:hist_scatter} but only for detections in these three bands. Colors of the galactic disk are again shown in green numbers as in Fig.\ref{fig:young_star_color_mag}. The median F275W $-$ F475W error is 0.27 mag and 0.31 mag for the tail and control region, respectively. The median F475W $-$ F814W errors are 0.01 mag in both regions. An evolutionary track of a star cluster, after an initial starburst, is derived from SB99, with the nebular emission added from Cloudy simulations and plotted. The specific track is the Geneva V00 track. The markers on the track indicate ages of 1, 3, 5, 10, 30, 50, 100, 200, 500 and 1000 Myr, starting from the blue end. A track with an intrinsic extinction of E(B-V) = 0.161 is also shown. The source histogram in each region, normalized to the region area, is again shown. For comparison, the Gaussian kernel density estimation~(KDE) of the histogram in Fig.~\ref{fig:hist_scatter} is also shown. 
    Again there is no evidence of active SF in the tail region.
    }
    \label{fig:hist_scatter_2}        
\end{figure*}

Source detection and aperture photometry were again performed in all bands with SExtractor, with an aperture radius of 0.5$''$. The F475W data were used as the detection.\footnote{If the F275W data were taken as the detection, far fewer sources are detected but our conclusion stays the same.}
Before the final color - color and color - magnitude diagrams, we filtered out spurious sources, weak sources, very red sources and foreground stars with the following steps. 
Bright foreground stars are excluded with magnitude cuts of $<$ 20.30 mag in F475W and $<$ 20.06 mag in F814W, using GSC2.
We also exclude sources with a color error greater than one magnitude. Color limits of F275W $-$ F475W $>$ 5 and F475W $-$ F814W > 3.5 were applied to exclude very red sources. Saturated stars and artifacts are also excluded by visual inspection. Many sources detected in F475W are not detected in F275W so only the F275W - F475W limits are shown for those sources.
In the end, there are 369 sources with 265 as upper limits in the control region and 38 sources with 18 as upper limits in the tail region .
The color - color diagram (F275W $-$ F475W versus F475W $-$ F814W) for all detected sources with and without upper limits are shown in Fig.~\ref{fig:hist_scatter} and Fig.~\ref{fig:hist_scatter_2}, respectively.
We also use 11 boxy regions (a length of 3.0$''$ $\times$ a width of 2.6$''$) along the galactic disk detected in the F275W image (Fig.~\ref{fig:composite}) to study the color of the galactic disk, which can be compared with the light and color profiles in Fig.~\ref{fig:eso_profile}.

\begin{figure*}
\vspace{-0.2cm}
    \includegraphics[width=0.65\textwidth]{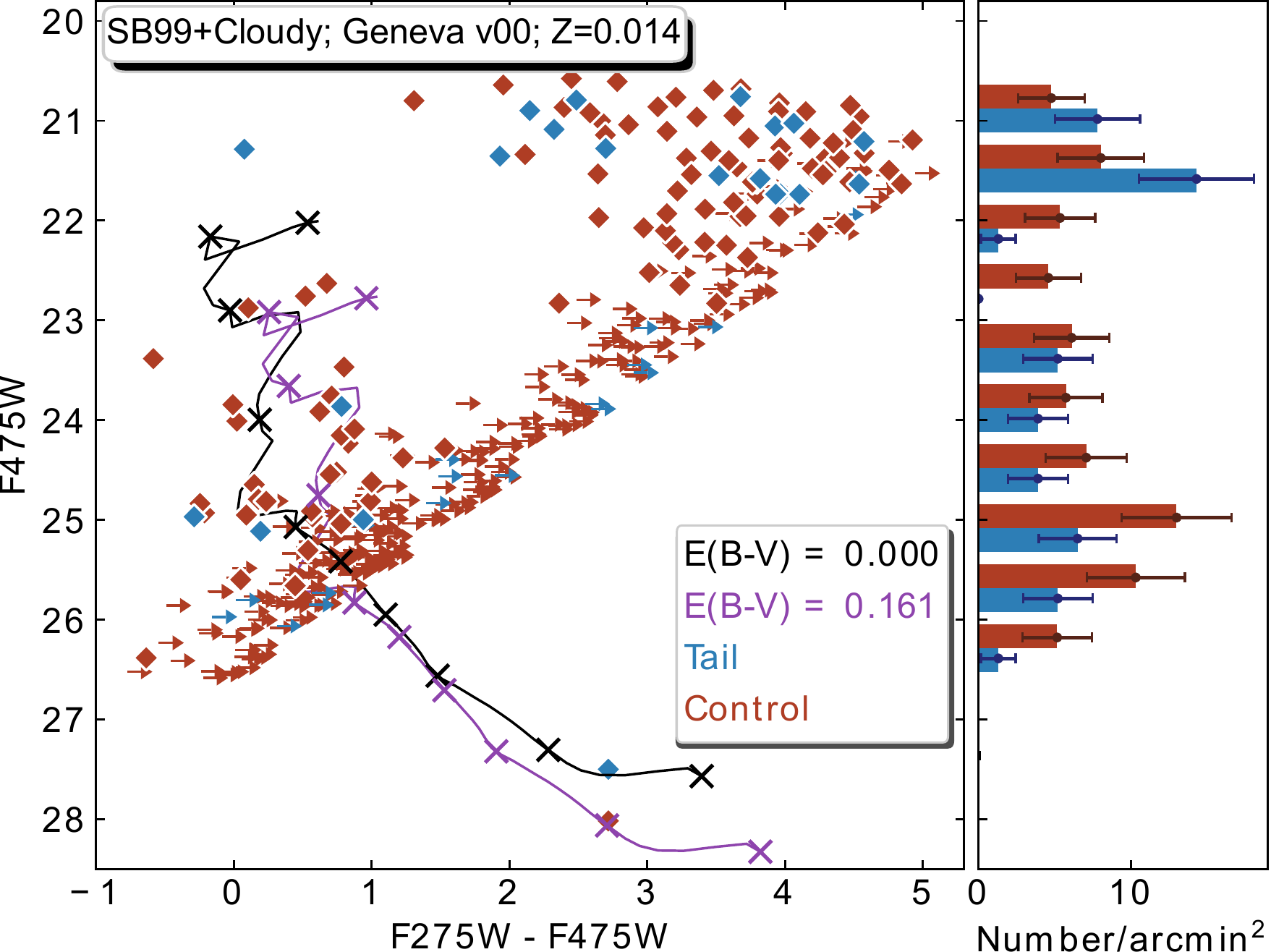}
    \vspace{-0.2cm}
    \caption{The F275W - F475W color --- F475W magnitude diagram for sources detected in F275W, F475W and F814W (diamonds) and sources only detected in F475W and F814W (arrows), in tail (blue) and control (red) regions. The SB99 tracks (for a total mass of $10^4$ M$_\odot$), with the nebular emission added from Cloudy simulations, are also shown with E(B-V) = 0 and 0.161. The markers on the track are same as in Fig. \ref{fig:hist_scatter_2}. A histogram of the source surface number densities in these two regions is shown on the right, which suggests no significant difference in the F475W magnitude distribution between sources in the tail and control regions. The somewhat higher number of bright sources in the tail region are almost all red ones (e.g., F275W - F475W $>$ 1 mag) and the only bright blue tail source is likely a Galactic star for its lack of H$\alpha$ emission and bright NIR emission from F160W. Combined with Fig.~\ref{fig:hist_scatter} and Fig.~\ref{fig:hist_scatter_2}, the source population in the tail region has no significant difference from that in the control region and there is no evidence for active SF in the tail region.
    }
    \label{fig:color_mag}        
\end{figure*}

To explore whether there is an excess of blue sources in the tail, we compare the color - color and color - magnitude diagrams for the sources in the tail and control regions. Particularly, histograms (the discrete form or the KDE form) for both colors and the F475W magnitude are compared. As shown in Fig.~\ref{fig:hist_scatter} and Fig.~\ref{fig:hist_scatter_2}, there is no evidence for any excess of blue sources (bright or faint) in the tail. We also show the color - magnitude diagram for sources in the tail and control regions in Fig.~\ref{fig:color_mag}. Again, there is no evidence for a significant excess of blue sources in the tail at any magnitude range.
Thus, there is no evidence of active SF in the stripped tail of \eso{}, which is consistent with the lack of compact H$\alpha$ sources in the tail \citep{2010ApJ...708..946S,2013ApJ...777..122Z}.

We also plot the SB99+Cloudy tracks, one without intrinsic extinction and another with an intrinsic extinction of E(B-V)=0.161 on the color - color and color - magnitude diagrams in Fig.~\ref{fig:hist_scatter_2} and Fig.~\ref{fig:color_mag} to show the expected positions of young star clusters. This particular intrinsic extinction value comes from the median intrinsic extinction of $A_{\rm V}$ = 0.5 mag derived for GASP clumps from \cite{2019MNRAS.482.4466P} and the assumed $R_{\rm V}$ value of 3.1.
Since our analysis suggests no excess of blue sources in the tail, we assume a F475W upper limit of 25.7 mag for any faint star clusters in the tail (Fig.~\ref{fig:color_mag}).
Based on the track in Fig.~\ref{fig:color_mag}, we can put an upper limit on the mass of young star clusters in the tail at  $\sim 2.1 \times 10^3$ M$_{\odot}$ at an age of 10 Myr when no intrinsic extinction is applied. For even younger star clusters, the upper mass limit will be smaller. For star clusters with an age of 100 Myr, the upper mass limit becomes $\sim 1.3 \times 10^4$ M$_{\odot}$. When intrinsic extinction is applied, the upper mass limit becomes $\sim 4.2 \times 10^3$ M$_{\odot}$ at an age of 10 Myr and $\sim 2.6 \times 10^4$ M$_{\odot}$ at an age of 100 Myr respectively.

%% file: CO_emission.tex
\section{CO emission from the galaxy and the tail}\label{sec:CO_emission}
The presence of the bright X-ray and H$\alpha$ emission over the center and the southern side of the disk, and its absence in the northern disk side, suggest that (diffuse) ISM has been displaced by ram pressure. It is thus of interest to inspect how the dense ISM component has been affected. As we will show in this section, about $5.5 \times 10^{9}$~M$_\odot$ of molecular gas was detected in total, distributed asymmetrically in the galaxy and the inner tail.

\subsection{Center of ESO~137-002}
\begin{figure} 
    \includegraphics[width=0.47\textwidth]{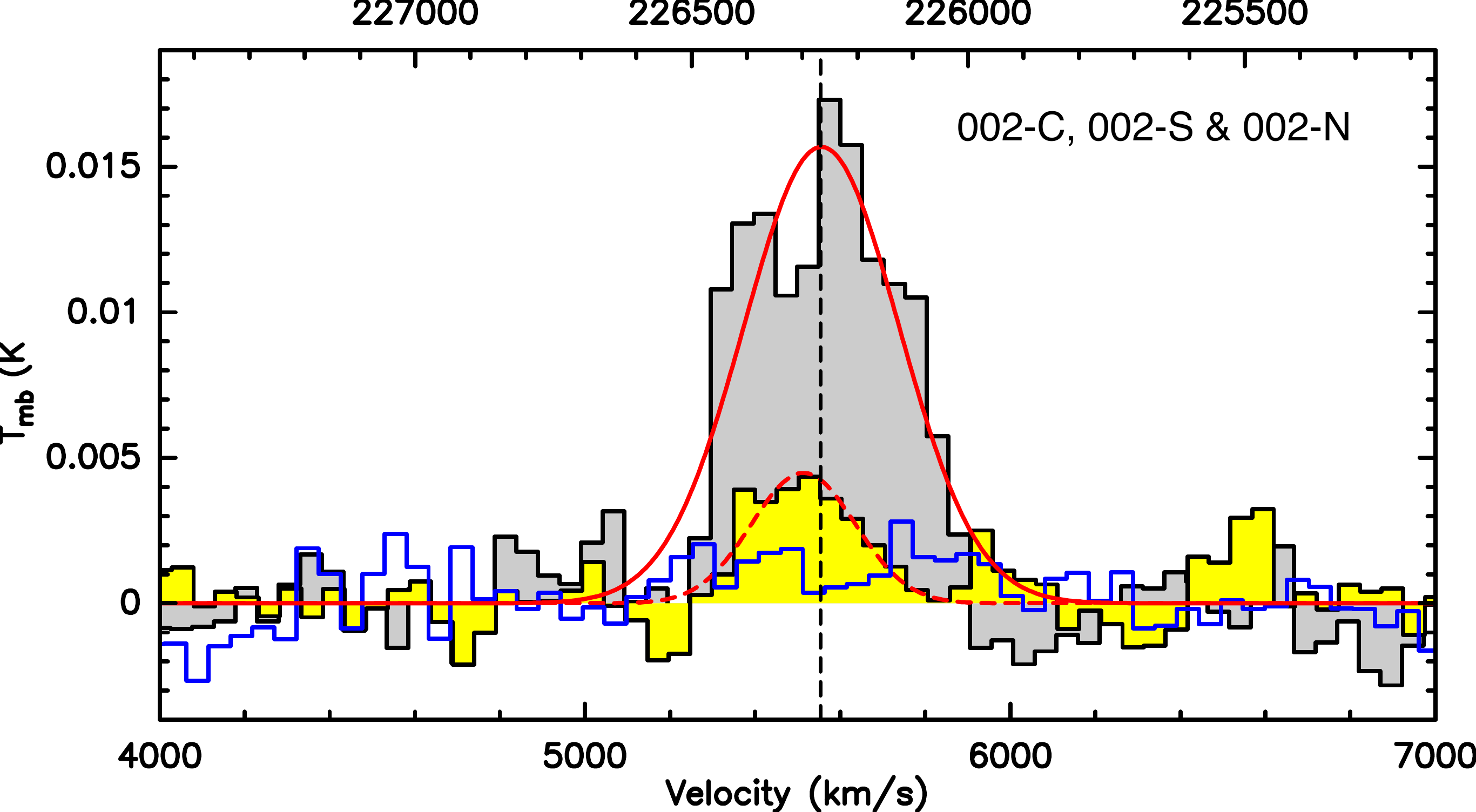}
    \vspace{-0.2cm}
    \caption{The CO(2-1) spectra of \eso{} in the central pointing 002-C (grey), the southern off-center position 002-S (yellow), and the symmetrical northern position 002-N (blue) (see Fig.~\ref{fig:region} for the positions of CO regions). Spectra are smoothed to 50.8~km\,s$^{-1}$ channels. Parameters of the Gaussian fits to the 002-C and 002-S lines are given in Table~\ref{tab:apexRes}. The central velocity 5554~km\,s$^{-1}$ of the fit in central point is shown with the dashed line, while the systemic radial velocity of \eso{} is 5585~km\,s$^{-1}$ (both assuming radio definition of the velocity and the latter velocity value corresponds to 5691~km\,s$^{-1}$ in the optical definition of the velocity).
    }
    \label{fig:apex_galaxy}
\end{figure}

We first searched for CO emission in the 002-C pointing covering the central part of the disk, where the gravitational restoring force is the strongest. CO(2-1) emission is strongly detected there (see Fig.~\ref{fig:apex_galaxy} and Table~\ref{tab:apexRes}). We calculate the CO luminosity from the standard relation of \citet{2005ARA&A..43..677S}
\begin{equation}\label{eq:CO}
L'_{\rm CO}= 3.5\times 10^7\, S_{\rm CO}\, \Delta\upsilon\, \nu_{\rm obs}^{-2}\, D_L^2\, (1+z)^{-3},
\end{equation}
where $L'_{\rm CO}$ is the CO(2-1) line luminosity in $\rm K\, km\, s^{-1}\, pc^2$, $S_{\rm CO}\, \Delta\upsilon$ is the CO velocity integrated line flux in $\rm Jy\, km\, s^{-1}$, $\nu_{\rm obs}$ is the CO(2-1) line observed frequency, and $D_L$ is the distance in Mpc. The CO luminosity in the 002-C pointing is $\sim 8.2\times 10^8$~$\rm K\, km\, s^{-1}\, pc^2$. Following
\begin{equation}
M_{\rm H_2}\, [M_\odot]= 5.4\, L'_{\rm CO(2-1)}\, [{\rm K\,km\,s^{-1}\,pc^2}],
\end{equation}
where we assume a standard Galactic CO/H$_2$ conversion factor $\alpha_{\rm CO}= 4.3~M_\odot/($K\,km\,s$^{-1}\,$pc$^2)$ \citep[e.g., ][]{2009AJ....137.4670L, 2012ARA&A..50..531K}, and a typical value for the CO(2-1)/CO(1-0) ratio of 0.8, the corresponding molecular gas mass in the 002-C region is about $4.4\times 10^9$ M$_\odot$. The formula includes a factor of 1.36 to account for the contribution of Helium. We assume a typical CO(3-2)/CO(1-0) ratio of 0.6.

\begin{table*}
    \begin{center}
    \caption{APEX results: The $1\sigma$ rms, the parameters of single Gaussian fits, the measured integrated intensity, and molecular gas mass are given for CO(2-1) emission detected in the observed positions. Temperatures are given in $T_{\rm mb}$ scale. First order (third order for 002-S) baselines were subtracted in the velocity range $4000 - 7200$~km\,s$^{-1}$. 
    }
    \label{tab:apexRes}
    \begin{tabular}{cccccccccc}
        \hline
        Source & Line & rms$^a$ & Velocity$^b$ & FWHM & $T_{\rm peak}$ & $I_{\rm CO, fit}$ & $L_{\rm CO}$ & $M_{\rm mol}$\\
        & & (mK) & (km\,s$^{-1}$) & (km\,s$^{-1}$) & (mK) & (K~km~s$^{-1}$) & ($10^8$~K~km~s$^{-1}$~pc$^2$) & ($10^8 M_\odot$)\\
        \hline
        002-C & CO(2-1) & 1.7 & $5659.3\pm 8.8$ & $429.1\pm 17.8$ & 15.7 & $7.2\pm 0.3$ & 8.2 & 44.3\\
        002-C-CO32 & CO(3-2) & 3.1 & $5687.9\pm 14.4$ & $443.0\pm 29.8$ & 16.7 & $7.9\pm 0.5$ & 4.21 & 30.2\\
        002-S & CO(2-1) & 1.4 & $5617.4\pm 23.1$ & $301.7\pm 49.3$ & 5.0 & $1.6\pm 0.2$  & 1.8 & 9.8\\
        002-S-CO32 & CO(3-2) & 1.7 & $5681.9\pm 25.2$ & $321.0\pm 56.0$ & 4.3 & $1.46\pm 0.23$ & 0.78 & 5.6\\
        002-N & CO(2-1) & 1.2 & $-$ & $\sim 300$ & $-$ & $<0.32^c$ & $<0.39^c$ & $<2.1^c$\\
        002-T & CO(2-1) & 0.9 & $5651.7\pm 21.3$ & $148.9\pm 39.7$ & 2.2 & $0.35\pm 0.09$ & 0.4 & 2.2\\
        \hline
    \end{tabular}
    \end{center}
    \vspace{-0.2cm}
     Note: $^\mathrm{(a)}$ in 25~km/s channels (or 27 km/s channels for CO(3-2) lines); $^\mathrm{(b)}$ heliocentric central velocity (optical definition);
     $^\mathrm{(c)}$ upper limit for 3$\sigma_{\rm rms}$
\end{table*}

A Gaussian fit to the 002-C CO(2-1) line (shown in Fig.~\ref{fig:apex_galaxy}) has a width of $\sim 430$~km\,s$^{-1}$ (Table~\ref{tab:apexRes}) which is consistent with \eso{}'s circular velocity of $\sim 200$~km\,s$^{-1}$ that is estimated from the galaxy's $K$-band luminosity and using the velocity-luminosity relation of \citet{2007ApJ...671..203C}.
While the CO line seems to have a classical two-horned profile caused by the disk rotation, it is asymmetric with (a) more pronounced low-velocity side and (b) strong peak in the middle of the line. The former is likely due to pointing the APEX 002-C aperture about 1.4~kpc southwards off the (optical) galaxy center where it encloses more gas from the southern disk side that rotates towards the observer, i.e. at smaller radial velocities. The latter may suggest the existence of a gas with a low radial velocity component relative to the galaxy, as the peak occurs at the galaxy's systemic radial velocity of 5691~km\,s$^{-1}$. Its origin is likely in a compact central source, such as a circumnuclear disk~\citep[e.g.,][]{1999A&A...344L..83I}. Many observations have shown that molecular gas is more centrally concentrated in galaxies with bars, suggesting that a bar can play an important role in gas-fueling toward the central region, thus feeding starbursts or AGN activities~\citep{2007PASJ...59..117K}. The average central gas surface density of barred spirals is known to be about three times higher than that of unbarred galaxies \citep{2005ApJ...632..217S}. \eso{} indeed has a bar as discussed in Section~\ref{sub:morph} and it hosts a Seyfert-2-like nucleus~\citep{2010ApJ...708..946S,2013ApJ...777..122Z}.

\subsection{Asymmetric CO distribution in the disk}
To cover the continuation of the bright X-ray and H$\alpha$ emission in the southern part of the disk, the 002-S {\em APEX} beam was pointed $\sim 7.8$~kpc south from the optical disk center, adjacent to the 002-C region (see Fig.~\ref{fig:region}). A rather strong CO(2-1) line is detected with a S/N~$\sim 12$
(see Fig.~\ref{fig:apex_galaxy}), defined as ${\rm S/N}= I_{\rm CO}/(\sigma_{\rm rms}\sqrt{\Delta\upsilon_{\rm CO}d\upsilon})$, i.e., a ratio of the integrated intensity and the noise integrated over the channels covered by the line, where $\Delta\upsilon_{\rm CO}$ is the channel width and $d\upsilon$ the integration range approximated by the FWHM of the line. The corresponding molecular gas mass is $\sim 1.0\times 10^9$ M$_\odot$. The linewidth measured from a Gaussian fit is $\sim 302$~km\,s$^{-1}$, by about 130~km\,s$^{-1}$ less than in the central 002-C pointing. The line is asymmetric, with more emission coming from the low-velocity side, which is consistent with detecting the gas that is rotating towards the observer. The high-velocity wing of the line exceeds the galaxy's systemic radial velocity, suggesting that the kinematics of the gas in the southern disk part may be altered by the effects of ram pressure. However, contamination by the secondary beam lobes that may register emission from the receding (northern) disk side is also possible.

We further observed the region 002-N located in the northern, upstream disk side, symmetrically to the southern 002-S pointing with respect to the optical center of the disk (see Fig.~\ref{fig:region}). No obvious line emission appears in the 002-N spectrum. From the measured rms noise $\sigma_{\rm rms}$ we calculate $3\sigma$ upper limit on the CO flux density
\begin{equation}
S_{\rm CO}< 3\sigma_{\rm rms}\, \sqrt{\Delta\upsilon_{\rm CO}\, d\upsilon},
\end{equation}
where $d\upsilon$ is the channel size
and $\Delta\upsilon_{\rm CO}$ is the mean FWHM linewidth. Assuming a linewidth of 300~km\,s$^{-1}$, the rms corresponds to a $3\sigma$ mass sensitivity of $\sim 2.1\times 10^8$ M$_\odot$. In the northern, windward part of the disk of \eso{}, there is thus at least 5-times less molecular gas than in its southern, downstream part. This indicates a strong asymmetry in the molecular component distribution in the disk, likely due to the effects of ram pressure stripping.

\subsection{Inner tail of ESO~137-002}

\begin{figure} 
    \includegraphics[width=0.47\textwidth]{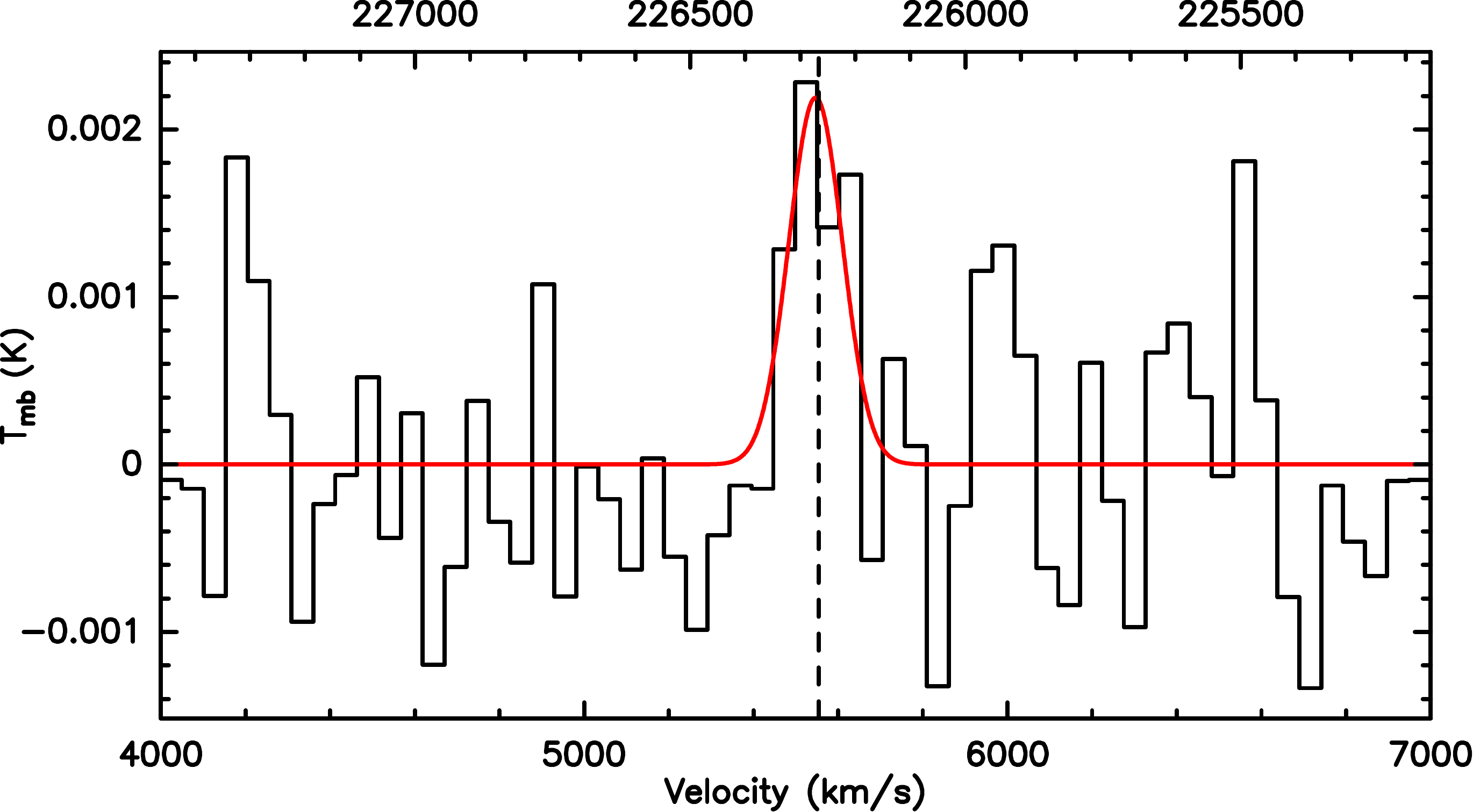}
    \vspace{-0.2cm}
    \caption{The CO(2-1) spectrum measured in the (inner) tail of \eso{} (002-T in Fig.~\ref{fig:region}). Spectrum is smoothed to $\sim 51$~km\,s$^{-1}$ channels. Parameters of the Gaussian fits are given in Table~\ref{tab:apexRes}. The central velocity of the line in the 002-C region is indicated with the dashed vertical line.
    }
    \label{fig:apex_tail}
\end{figure}

We also observed the inner tail region (002-T) (see Fig.~\ref{fig:region}). The integration was deep to reach an rms sensitivity of $\sim 0.7$~mK in 51~km\,s$^{-1}$ channels. Fig.~\ref{fig:apex_tail} shows the spectrum where the CO line is detected with an `integrated' S/N > 6.
The FWHM of the line is considerably smaller than in the 002-S region - only $\sim 149$~km\,s$^{-1}$ (by about a factor of 2.2). The line luminosity corresponds to an H$_2$ mass of $\sim 2.2\times 10^8$ M$_\odot$, and the integrated intensity corresponds to a column density of $\sim$ 2M$_\odot$\,pc$^{-2}$ averaged over the beam. At a higher resolution, the emission would likely mostly split into a set of compact regions with higher column densities. The observed linewidth of $\sim 150$~km\,s$^{-1}$ could correspond to a superposition of many molecular clouds with typical values of the velocity dispersion of $\sim 5-15$~km\,s$^{-1}$. Dust features, which are presumably good tracers of molecular gas, show mostly compact morphologies in the \hst{} image. At the same time, previous observations of other RPS galaxies indicated from comparison of interferometric and single-dish fluxes that there might be an important extended component of the CO emission on top of a rich distribution of compact CO sources~(e.g. \citet{2019ApJ...883..145J} in ESO~137-001 or \citet{2020ApJ...889....9M} in JW100). It is thus possible that in ESO~137-002, a certain fraction of the molecular gas is also in the form of a more diffuse component. However, these observations suggested that the fraction of the extended emission increases with the distance along the tail and is low in the main galaxy and the inner parts of the RPS tails.

In another Norma cluster galaxy, ESO~137-001, large amounts of molecular gas were detected in the tail. Since \eso{} is stripped (and observed) edge-on, the tail overlies the disk, before it extends from the galaxy. In a face-on stripping configuration, the tail is extra-planar. The 002-T region thus should be compared to the 001-tailB region in ESO~137-001, where about 3-times more H$_2$ was detected~\citep{2014ApJ...792...11J}. 
The CO line is centered at a velocity close to the galaxy's systemic velocity ($\Delta v\sim 40$~km/s). This may indicate that a substantial fraction of the orbital velocity of the galaxy is happening in the plane of the sky, and the stripped gas has an important tangential velocity component.

\subsection{CO(3-2) emission and CO(3-2)-to-CO(2-1) line ratios}
\begin{figure} 
    \includegraphics[width=0.47\textwidth]{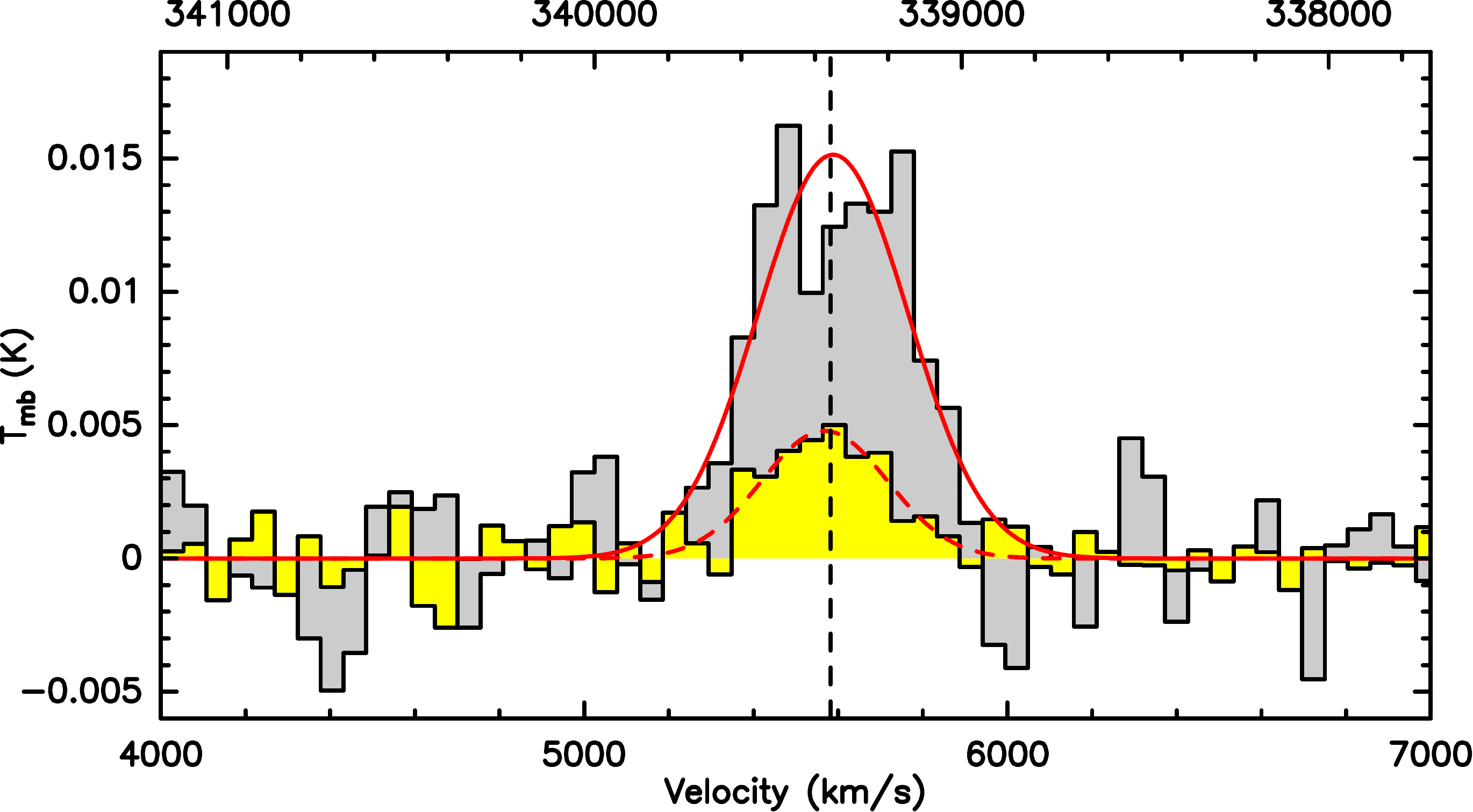}
    \vspace{-0.2cm}
    \caption{
    The CO(3-2) spectra measured in the central (grey) and southern (yellow) positions over \eso{} (see positions in Fig.~\ref{fig:region}). Spectra are smoothed to $\sim 50$~km\,s$^{-1}$ channels. Parameters of the Gaussian fits are given in Table~\ref{tab:apexRes}. The central velocity of the line in the 002-C region is indicated with the dashed vertical line. The spectrum of the central position is double-peaked with a velocity separation of $\sim$ 300 km/s.
    }
    \label{fig:apex_co32}
\end{figure}

The central and southern regions of the ESO~137-002 disk were further observed in CO(3-2). The locations of the two pointings were slightly offset with respect to the CO(2-1) regions 002-C and 002-S in order to better cover the asymmetric X-ray and H$\alpha$ emission with the smaller CO(3-2) beams (see Table~\ref{tab:apexObs} and Fig.~\ref{fig:region}). The corresponding CO luminosities and H$_2$ masses are given in Table~\ref{tab:apexRes}. In the central region the CO(3-2) line is double-horned
with some emission also at the central line velocity, similarly to the CO(2-1) 002-C line profile (see Fig.~\ref{fig:apex_co32}). The CO(3-2) lines in the central and southern regions have slightly larger FWHM than the corresponding CO(2-1) lines. The difference, however, is minimum and well within the error of the Gaussian line fits. Other factors may play a role, such as noise, differences in beam sizes, beam offsets, and possible source extension.

Following Eq.~\ref{eq:CO}, the line luminosity ratios, $r_{32}= L'_{\rm CO(3-2)}/L'_{\rm CO(2-1)}$ in the two regions are 0.51 and 0.43, not corrected for the different beam sizes ($\sim 27''$ vs. $17''$). The galaxy is edge-on and it can be expected that most of the molecular gas is concentrated in the disk plane. Also, the ram pressure operating on the galaxy is nearly edge-on. The difference in the beam sizes in the direction perpendicular to the disk-plane thus does not play much
role and it is mostly the difference along the disk plane that matters. We thus simply apply a 1D geometrical beam correction factor of $26.7''/17.8''= 1.5$ that increases the $r_{32}$ line ratios in the 002-C-CO32 and 002-S-CO32 regions to 0.77 and 0.65, respectively.
A more precise way for calculating a beam correction factor would be to follow the assumption that the IR dust emission is a good tracer of the cold molecular gas and measure the flux within different apertures in the IR images. But this approach is beyond the scope of this paper.

In low-redshift normal star-forming galaxies, typical mean values of the CO line ratios are $r_{31}= 0.55-0.61$~\citep{2010ApJ...724.1336M, 2017ApJS..233...22S}, and $r_{21}=0.79-0.81$ \citep{2009AJ....137.4670L,2017ApJS..233...22S}. This translates to the typical $r_{32}$ ratio of $\sim 0.7$. Our measurements in the two regions in ESO~137-002 are thus consistent within the uncertainties with the typical values in local star-forming galaxies. The line luminosity ratio can be interpreted as an indicator of the gas density. The critical density of the CO(3-2) transition calculated under the optically thin assumption is $3.6\times 10^4$~cm$^{-3}$. The $r_{32}$ line ratio may thus be considered a proxy for the ratio of relatively dense to more diffuse molecular gas. Many studies have searched in galaxies for correlations between the CO line ratios and star formation rates or star formation efficiencies (SFE), finding that the $r_{31}$ ratio tends to increase with SFE, suggesting that galaxies with a higher fraction of dense molecular gas tend to have higher SFE~\citep[e.g.,][]{2020ApJ...889..103L}. Despite a large scatter in the correlation~\citep[see Fig. 3 in ][]{2020ApJ...889..103L}, it is clear that the values of the line ratios measured in the two regions in ESO~137-002 would in a normal star-forming galaxy correspond to a larger SFE. In other words, despite its large amounts of molecular gas and the relatively large fraction of the denser molecular gas traced by the CO(3-2) emission, ESO~137-002 has a rather low SFE: ${\rm SFE}= {\rm SFR}/M_{H_2}\approx 0.89$~M$_\odot\, {\rm yr}^{-1} / 5.5\times 10^9$~M$_\odot\sim 1.6\times 10^{-10}$~yr$^{-1}$. The effects of ram pressure of the surrounding ICM, as well as the active nucleus (i.e. a source of X-rays) of the galaxy, may be responsible for its measured low SFE.

%% file: discuss.tex
\section{Discussion}\label{sec:discussion}

\subsection{Star formation in the galaxy}\label{subsec:SF_gal}

The total FIR luminosity of the galaxy was derived from the {\em Herschel} data. We used the {\em Herschel} source catalog, particularly the PACS Point Source Catalog and the SPIRE Point Source Catalog. \eso{} is significantly detected in all six {\em Herschel} bands. We used the python code MBB\_EMCEE to fit modified blackbodies to photometry data using an affine invariant Markov chain Monte Carlo (MCMC) method, with the {\em Herschel} passband response folded~\citep{2014ApJ...780...75D}. Assuming that all dust grains share a single temperature $T_{\rm d}$, that the dust distribution is optically thin, and neglecting any power-law component towards shorter wavelengths, the fit results in a temperature of $T_{\rm d}$ = (27.8$\pm$0.2) K, a luminosity $L_{\rm 8-1000 \ \mu m}$ = (1.22 $\pm$ 0.02) $\times$ 10$^{10}$ L$_{\odot}$, and a dust mass of $M_{\rm d}$ = (6.7 $\pm$ 0.3) $\times 10^{6}$ M$_{\odot}$ for $\beta$ = 1.5.
For $\beta$ = 2, $T_{\rm d}$ = (24.7$\pm$0.2) K, $L_{\rm 8-1000 \ \mu m}$ = (1.21 $\pm$ 0.02) $\times$ 10$^{10}$ L$_{\odot}$ and $M_{\rm d}$ = (1.2 $\pm$ 0.1) $\times 10^{7}$ M$_{\odot}$.
We also performed a fit using a free $\beta$ parameter, which results in $T_{\rm d}$ = 26.6$^{+0.9}_{-0.7}$ K, $L_{\rm 8-1000 \ \mu m}$ = (1.22 $\pm$ 0.02) $\times$ 10$^{10}$ L$_{\odot}$, $M_{\rm d}$ = 8.7 $^{+0.9}_{-1.6} \times 10^{6}$ M$_{\odot}$, and $\beta$ = 1.69$^{+0.09}_{-0.14}$.
With the measured FIR and CO luminosities, we can put \eso{} on the well-known CO - FIR correlation \citep[e.g.,][]{2005ARA&A..43..677S} and its position is consistent with normal spirals on the relation.

The total SFR of \eso{} is 1.08 M$_{\odot}$/yr, from the {\em Galex} NUV flux density and the total {\em Herschel} FIR luminosity from the relation of \citet{2011ApJ...741..124H}. If using the {\em WISE} 22 $\mu$m flux density and the relation from \citet{2013ApJ...774...62L}, the estimated total SFR is 0.80 M$_{\odot}$/yr. The Kroupa IMF is assumed in both cases. The \citet{2013ApJ...774...62L} work assumed the Salpeter IMF so we multiply its SFR relation by 0.62 to convert to the Kroupa IMF. For the total amount of molecular gas in the galaxy, $\sim 5.5 \times 10^{9}$ M$_{\odot}$, the molecular gas depletion time is $\sim$ 5.9 Gyr.

\eso{} is $\sim$ 5 times more massive than ESO~137-001 but with a comparable SF activity. The specific SFR of \eso{} is $\sim$ 7 times less than that of ESO~137-001. This is also consistent with their different $W1 - W4$ colors, 4.25 for \eso{} vs. 6.99 for ESO~137-001, as $W1 - W4$ is a proxy for the specific SFR. Because of its prominent bulge and higher surface brightness in optical and NIR, \eso{} is in fact a more compact galaxy than ESO~137-001.

The upstream part of the galaxy is nearly dust free and gas free~\citep{2010ApJ...708..946S,2013ApJ...777..122Z} so the current SF is mainly around the nucleus and the downstream. Fig. \ref{fig:hist_scatter} and Fig. \ref{fig:hist_scatter_2} show that there is little SF in the tail. However, the disk hosts some blue star clusters with recent SF most likely triggered by ram pressure (Fig. \ref{fig:young_star_region}). The color $-$ color diagram~(Fig. \ref{fig:young_star_color_mag}) shows that these star clusters are indeed young, although the total SFR of these young star clusters is very small compared with the total SFR of the galaxy.

\subsection{Star formation efficiency in the tail}
\label{subsec:SF_tail}

\cite{2010ApJ...708..946S} presented the H$\alpha$ data of \eso{}, from the narrow-band imaging observations with {\em SOAR}. While the {\em SOAR} H$\alpha$ observations are not deep (three exposures of 800 sec each), the data were taken under the excellent seeing condition (0.45$''$ - 0.65$''$). The continuum data were taken from another narrow-band filter centered at $\sim$ 87\AA{} blueward the H$\alpha$ filter. Similar to what we did for ESO~137-001 with the {\em SOAR} data in \cite{2007ApJ...671..190S}, the source colors from the two narrow band filters are examined to select emission-line sources in the X-ray/H$\alpha$ tail of \eso{}. None is detected and the 3-$\sigma$ upper limit of the H$\alpha$ luminosity on the compact sources is 7$\times10^{37}$ erg s$^{-1}$, within an aperture radius of 0.6$''$ and assuming \NII\ $\lambda$6584 / H$\alpha$ = 0.2 and \NII\ $\lambda$6548 / \NII\ $\lambda$6584 = 1/3.
This limit is 4 times smaller than the median H$\alpha$ luminosity of 4$\times10^{38}$ erg s$^{-1}$ for over 500 H$\alpha$ clumps in the tails of GASP galaxies \citep{2019MNRAS.482.4466P}, even after the correction for the intrinsic extinction (with the median $A_{\rm V}$ of 0.5 mag from \citealt{2019MNRAS.482.4466P}). This limit is also smaller than $\sim$ 90\% of \hii{} regions detected by {\em SOAR} in ESO~137-001's tail \citep{2007ApJ...671..190S}. 
With the SFR-H$\alpha$ calibration from \cite{2011ApJ...741..124H} for the Kroupa IMF, the corresponding SFR limit is 3.9$\times10^{-4} \text{M}_{\odot} \text{yr}^{-1}$.
Thus, the SF in \eso{}'s tail is weak at most.

The \hst{} data also do not reveal any significant young star clusters in the tail, even with its superior angular resolution. Young star clusters with ages of less than 10 Myr have strong H$\alpha$ emission and will be detected as \hii{} regions. Observationally, compact UV emission that traces young star clusters correlates well with compact H$\alpha$ emission that traces \hii{} regions \citep[e.g.][]{2018MNRAS.479.4126G}.
With the \hst{} data, we set an upper limit of mass of M$_{\star}$ $\approx$ 4.2 $\times 10^3$ M$_{\odot}$~(with $A_{\rm V}$ = 0.5 mag intrinsic extinction) at an age of 10 Myr. The corresponding upper limit on the SFR in the tail is,
\begin{equation}
    \textrm{SFR}_{\textrm{tail}} < \frac{4.2 \times 10^3 \textrm{M}_{\odot}}{1 \times 10^7 \textrm{yr}} < 4.2 \times 10^{-4} \text{M}_{\odot} \text{yr}^{-1}
\end{equation}

Note that the resulting SFR limit would be smaller if using an age of 100 Myr and assuming no intrinsic extinction.
The inner tail~(002-T) of \eso{} from the \apex{} data corresponds $\sim 2.2 \times$ 10$^8$ M$_{\odot}$ of H$_2$. With the larger SFR limit derived earlier, we estimate the SFE as, 
\begin{equation}
    \text{SFE}_{\text{tail}} < \frac{4.2 \times 10^{-4} \text{M}_{\odot} \text{yr}^{-1}}{2.2 \times 10^8 \text{M}{_\odot}} < 1.9 \times 10^{-12} \text{yr}^{-1}
\end{equation}

Taking the reciprocal of the SFE, the gas depletion time in 002-T is 526 Gyr. The SFE upper limit is more than 19 times lower compared to the ESO 137-001 tail~(SFE $\approx$ 3.6 $\times$ 10$^{-11}$ yr$^{-1}$)~\citep{2014ApJ...792...11J}. It is also smaller than the SFE in D100's tail ($\sim 6.0 \times 10^{-12}$ yr$^{-1}$).
The SFE typically falls with the distance from the main body to the outer tail as in the case of other RPS galaxies \citep[e.g.,][]{2012A&A...545A.142B,2014ApJ...792...11J}. 

\subsection{Stripping history of ESO 137-002}\label{sub:stripping_history}

We can compare the amount of gas at different phases (cold molecular, warm ionized and X-ray emitting) detected in the galaxy and its tail. 

\begin{itemize}
    \item Cold molecular gas \\
    We detected $\sim 5.5 \times 10^{9}$ M$_{\odot}$ of the cold molecular gas from the galaxy and at least $\sim 2\times10^{8}$ M$_{\odot}$ from the tail. Deeper observations plus wider coverage of the tail may reveal more cold molecular gas there. For the total stellar mass and the sSFR of \eso{}, the cold gas fraction of the galaxy is $\sim$ 20\% from the scaling relations by \cite{2014A&A...564A..66B} and \cite{2018MNRAS.476..875C}, although the big scatter can change the result by a factor of several. This fraction translates to $\sim 7\times10^{9}$ M$_{\odot}$ of the cold gas initially in the galaxy. Thus, \eso{} may still retain most of its initial cold ISM, especially since the cold atomic gas is not included yet for the lack of sensitive HI data. This would suggest that the stripping in \eso{} is likely still in the early stage, in term of the ISM content. On the other hand, the distribution of the ISM in the galaxy may have been significantly modified, as shown also by the dust distribution. 
    
    \item Warm ionized gas \\
    \cite{2013ApJ...777..122Z} gave a total mass of $\sim 1.9 \times 10^7$ M$_\odot$ for the H$\alpha$ emitting gas in the tail, if a filling factor of 0.2\% is adopted. Even if the filling factor is 10 times higher, the mass only increases by 3.2 times. The mass of the warm ionized gas in the galaxy is hard to estimate, for the contribution from star forming regions and intrinsic extinction, but it is unlikely to be comparable to the mass of the cold molecular gas.
    
    \item Hot X-ray emitting gas \\
    \cite{2013ApJ...777..122Z} estimated a total mass of $\sim 2 \times 10^8$ M$_\odot$ for the X-ray emitting gas in the tail, if the filling factor is close to one. The amount of the X-ray emitting gas in the galaxy is less due to its smaller volume.
\end{itemize}

Beside the total gas content, this work also reveals a strong asymmetry in the distribution of the CO gas in the disk, similar to that of the H$\alpha$ and X-ray emitting gas. While the galaxy may still retain a significant portion of its initial cold ISM, the ISM distribution has been significantly modified by ram pressure and narrow stripping tails have been formed behind the galaxy.

\begin{figure*}
    \vspace{-0.3cm}
    \centerline{\includegraphics[width=0.75\textwidth]{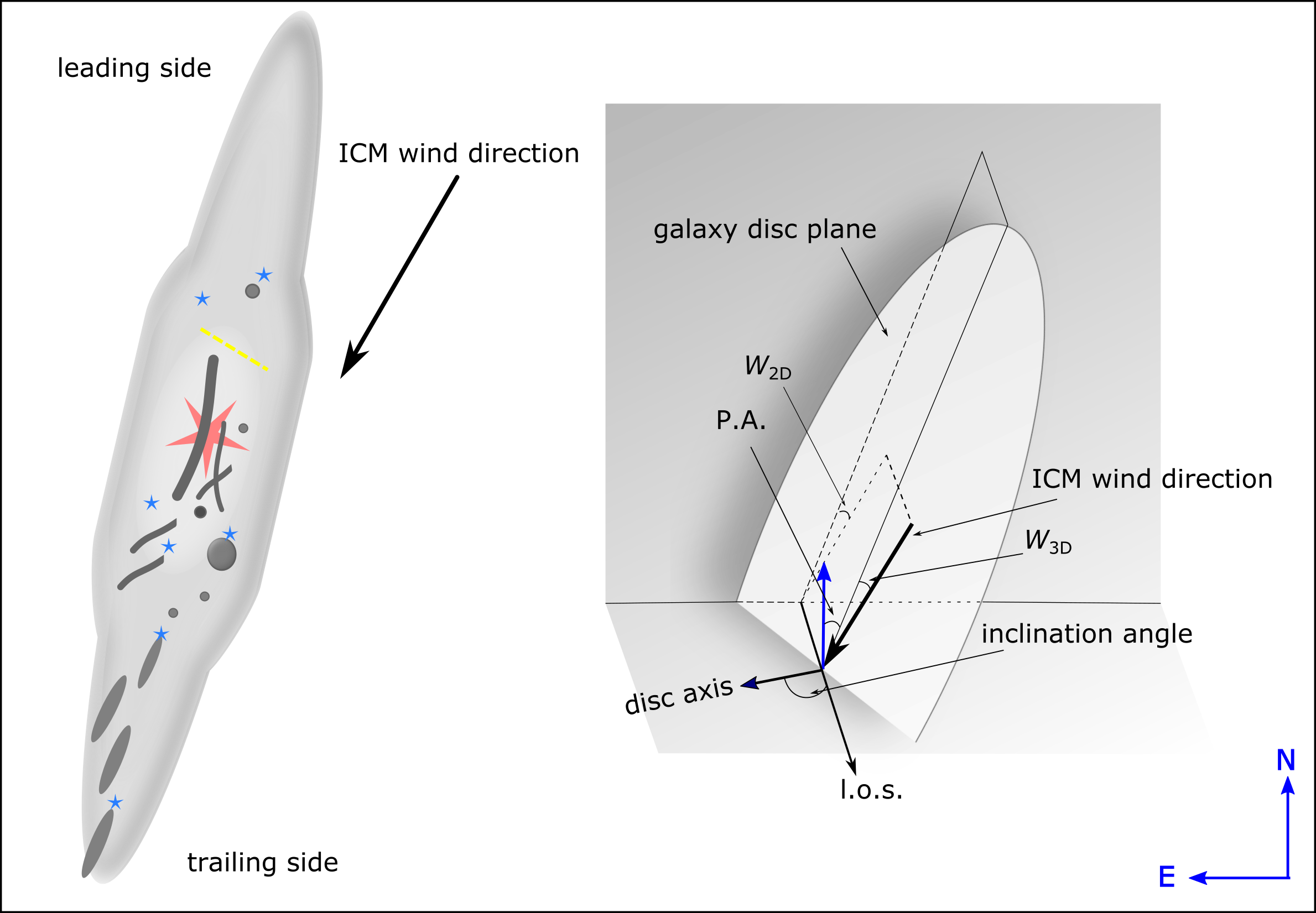}}
    \caption{
    {\em Left}: A cartoon representation of \eso{}, currently undergoing ram-pressure stripping. The ICM wind is projected to the galaxy from NW, dragging dust clouds and forming filaments at $\sim20^{\circ}$ to the east from the south. It is likely that the ablated dust clouds are located between the bright galactic disk and us to make the dust features more significant in the optical. This is consistent with the galaxy falling to the cluster with a redshift (or moving away from us). The model depicts dark clouds, along with star clusters in blue (some at the tip of ablated dust clouds), and a strong AGN (in red) at the nuclear position, plus a yellow stripping front separating the upstream and downstream regions. Note that the stripping front is defined by the H$\alpha$ and X-ray emission front, while stripping proceeds throughout the disk because the ICM wind is tilted relative to the disk plane ($\sim20^{\circ}$) and the ISM is porous.
    {\em Right}: A 2D representation of wind angle projected on the plane of sky. The direction of the ICM wind is nearly edge-on along the galactic disk. \eso{} likely has some line-of-sight motion with cluster such that the 3D disk-wind angle ($W_{\rm 3D}$) is less than the 2D projected disk-wind angle ($W_{\rm 2D}$).
    On the other hand, the asymmetric dust distribution in the galaxy suggests that the angle between the line-of-sight and the ICM wind cannot be too small (or too different from 90 deg).
    }
    \label{fig:eso_cartoon}
\end{figure*}

As discussed in \cite{2010ApJ...708..946S} and \cite{2013ApJ...777..122Z}, \eso{} is experiencing a near edge-on stripping. This is also supported by the \hst\ images of dust filaments in the disk, mostly aligned with the disk plane. The positions of young star clusters and their associated dust clouds also suggest near edge-on stripping. On the other hand, the orientations of most dust filaments, the H$\alpha$ and X-ray tails suggest a small projected wind angle ($W_{\rm 2D}$) of $\sim$ 20$^{\circ}$ between the ICM wind direction and the disk plane~(Fig.~\ref{fig:eso_cartoon}). Since much of the wind is projected in the plane of sky, it creates vertical distribution of the ISM.
\eso{} also has a significant line-of-sight motion with the cluster, so there must be a wind component perpendicular to the plane of the sky. For such case, the 3D disk-wind angle ($W_{\rm 3D}$) between the galactic disk and the ICM wind direction is less than  20$^{\circ}$~(using the technique of \citealt{2011AJ....141..164A}). This low inclined angle likely has a contribution to pushing the disk ISM from the leading side.
Simulations done on near edge-on stripping cases show the low efficiency of stripping \citep[e.g,][]{2001ApJ...561..708V,2006MNRAS.369..567R}.
It is also noted that the secondary H$\alpha$ tail also has the same orientation with respect to the disk plane \citep{2013ApJ...777..122Z}. 

The rich dust features revealed by \hst, including filaments and large clouds, likely disclose the distribution of the cold ISM clouds experiencing ram pressure.
The dust lane exhibits numerous dust threads with a typical length of $\sim$ 0.8 kpc and width of $\sim$ 0.1 kpc. NGC~4921~\citep{2015AJ....150...59K}, another galaxy undergoing near edge-on stripping in the Coma cluster, but with face-on view ($i \lesssim 25^{\circ}$), also reveal similar dust features indicating ongoing RPS.
The $\sim$ 20 kpc dust front on the leading (NW) side of NGC~4921 shows different types of filamentary structures with 0.5 - 1 kpc in length and 0.1 - 0.2 kpc in width, approximately in the wind direction.
Similar dust cloud filaments were also observed in NGC~4402 and NGC~4522~\citep{2014AJ....147...63A,2016AJ....152...32A}, mostly aligned with the projected wind direction. 
\citet{2015AJ....150...59K} proposed that dust clouds with different shapes reflect spatial variations in ram pressure acceleration, due principally to variations in gas density, leading to the partial decoupling of the densest clouds. 
The complex morphologies suggest the interplay of magnetic field, galactic rotation and turbulence.  

As shown by e.g., \citet{2014MNRAS.438..444B}, indeed edge-on stripping mostly compresses the ISM instead of removing them from the galaxy. This compression can result in enhanced SF in the galaxy and the SF can continue for $\sim$ 1 Gyr after the pericentre passage.
Since the colors of these young star clusters spread over a range (Fig.~\ref{fig:young_star_color_mag}), it is possible that the RPS triggered SF proceeds with different timescales in different clouds so there is not a single burst-like event in the disk and some clouds do not have associated star clusters at the moment, besides the likely different intrinsic extinction for different clouds. As the ages of these young star clusters are $\sim$ several Myr - 200 Myr, \eso{} would have moved for $\sim$ 200 kpc in the sky if its velocity in the plane of the sky is 1000 km/s. Thus, the RPS triggered SF likely started when the galaxy began to plunge into the central, dense part of the cluster ICM.

\eso{} presents a great case of an edge-on galaxy experiencing a nearly edge-on ram pressure wind with a large component in the plane of the sky. It is especially useful for showing the vertical ISM behavior during the strong, nearly edge-on stripping, and for getting a nearly edge-on view of the stripping features.
There are other galaxies undergoing edge-on stripping, but many are viewed face-on, e.g., NGC~4921, NGC~4654 \citep{2021A&A...645A.111L}, NGC~4501 \citep{2008A&A...483...89V}, and others are probably not only affected by RPS, e.g., UGC~6697 \citep{2017A&A...606A..83C}.
\eso{} thus renders a perfect prototype for studying the edge-on stripping because of the favorable combination of 3 angles --- nearly edge-on 3D disk-wind angle, disk viewed nearly edge-on, and much of the ICM wind 
projected in the plane of sky.

\subsection{RPS galaxies with little SF in the stripped tail}

\eso{} is a great example of RPS galaxies with little SF in the stripped ISM. It has been known that active SF is not always present in the stripped tails, e.g., some galaxies in the Coma and A1367 samples of RPS galaxies selected by H$\alpha$ diffuse tail by \cite{2010AJ....140.1814Y,2017ApJ...839...65Y} and the Virgo sample studied by \cite{2012A&A...545A.142B}.
D100~\citep{2007ApJ...660.1209Y,2017ApJ...839..114J}, another RPS galaxy with a remarkably long and narrow H$\alpha$ tail ($\sim$ 60 kpc) also has weak SF in the tail. The total SFR in the tail is found to be $\sim$ 6 $\times 10^{-3}$ M$_{\odot}$ yr$^{-1}$ \citep{2019ApJ...870...63C}.
At least 1/3 of RPS galaxies in \cite{2010AJ....140.1814Y,2017ApJ...839...65Y} have little to weak SF in their stripped tails, e.g., GMP~3071, RB~219, IC~3949, NGC~4853, CGCG~097-073 and CGCG~097-079.
\citet{2016A&A...587A..68B} studied NGC 4569, the most massive spiral galaxy in the Virgo cluster, and revealed the presence of a diffuse gas tail but without any associated young stellar components. 

Because of the lack of active SF in the stripped medium, the diffuse gas tails of the above RPS galaxies can be faint and missed in the optical and UV imaging. As described in \citet{2017ApJ...844...48P}, the key selection criterion of the GASP sample is the stripping signatures shown in the $B$-band morphologies of these galaxies. The majority of GASP galaxies shows bright star-forming clumps in the disk and/or along the tails.
The fraction of diffuse gas (not star forming) in the tails \footnote{The diffuse gas in the tails is different from the concept ``diffuse ionized gas (DIG)''. DIG is a warm ($\sim 10^4$ K) and low density ($\sim 0.1$ cm$^{-3}$) gas phase 
in the ISM of galaxies (see \citealt{2009RvMP...81..969H} for a review). In the stripping tails, while the diffuse gas surrounding the \hii{} regions may partly include the DIG, 
the diffuse gas far away from the \hii{} regions is most likely the result of mixing between the stripped ISM and the ICM.} 
may increase for the galaxies with significant gaseous tail features.
\cite{2019MNRAS.482.4466P} studied 16 GASP galaxies with clear stripping tails beyond the galaxy boundary and found that SF contributes to 64\% --- 94\% of the H$\alpha$ emission in the tails,
which is considered as the dominating ionization mechanism there.
By separating the diffuse gas and H$\alpha$ clumps (the places where SF occurs) with the H$\alpha$ surface brightness map, they found that, only in four galaxies, the diffuse ionized gas can contribute over 60\% of the H$\alpha$ emission in the tails.
Note that \cite{2019MNRAS.482.4466P} used the stellar continuum surface brightness to define the galaxy boundary.
Adopting $D_{25}$ (the isophotal level of 25 mag/arcsec$^2$ in the $B$-band) as a boundary of the galaxy main body, we also performed an independent inspection on the released H$\alpha$ surface brightness maps of 114 GASP galaxies and estimated the H$\alpha$ flux fraction of the diffuse gas in their tail regions beyond $D_{25}$.
For 12 galaxies with clear tail features beyond $D_{25}$, we apply SExtractor to select \hii{} region candidates from the H$\alpha$ surface brightness map by requesting point-like sources (CLASS$_{-}$STAR $>$ 0.9) with a low ellipticity ($e <$ 0.2) \citep[e.g.,][]{2016MNRAS.455.2028F}. Four galaxies in the full sample of 114 GASP galaxies have a fraction of diffuse gas larger than 50\% in the tail.
Only one of these 114 galaxies has a fraction of diffuse gas larger than 60\%, JO204 ($\sim$ 75\%) and its H$\alpha$ tail is only detected to $\sim$ 32 kpc from the nucleus, while the galaxy is $\sim$ 30\% more luminous than \eso{} at the {\em WISE} W1 band.
Thus, indeed the GASP sample mainly covers the RPS galaxies with active SF downstream in the stripped tails.
On the contrary, the H$\alpha$ flux fraction of the diffuse gas in \eso{}'s tail is over 95\% for the lack of \ion{H}{II} regions.
The fraction of diffuse gas that we discussed above is estimated from the H$\alpha$ surface brightness map, which may be underestimated, since the exact ionization mechanism of the H$\alpha$ clumps and \hii{} region candidates need to be considered with more information (e.g. emission-line ratios, see \cite{2021ApJ...907...22T} as an example). Multi-wavelength imaging and IFU observations can provide further information to effectively distinguish the star-forming from diffuse gas and identify the stripping tails undergoing different evolutionary states.

It is still unclear what determines the SF efficiency in the stripped tails. Little to weak SF in the stripped tails may happen at the early, late or even intermediate stage of stripping.
Considering the whole evolution process of the stripped gas tails, the RPS galaxies with/without star-forming trails are both important sub-groups of the full population of the RPS galaxies. 
If selected by SF tails behind the galaxy in the optical or UV (e.g., surveys like GASP or using the {\em Galex} data), \eso{} and NGC~4569 will not be considered as a ``jellyfish'' galaxy.
Future deep and wide-field surveys in H$\alpha$, HI, radio continuum and X-rays, insensitive to SF, will help to build unbiased sample of RPS galaxies and further explore how the stripped ISM mixing with the hot ICM and its evolution path.

%% file: conclude.tex
\section{Conclusions}\label{sec:conclusion}

We present detailed analysis of \eso{}, a large edge-on spiral with a boxy bulge currently undergoing nearly edge-on RPS, with the \hst{} and the \apex{} data.
The main results of this paper are:

\begin{enumerate}

\item The galaxy has an undisturbed stellar body but a strongly disturbed ISM distribution, as shown by the H$\alpha$ emission and the dust distribution (Fig.~\ref{fig:composite} \& Fig.~\ref{fig:galfit_unsharp}).
The upstream of \eso{} is nearly dust-free while the downstream is full of rich dust features, suggesting RPS has nearly cleared the north half of the galaxy and still continue to strip around the nucleus and the southern part.
This asymmetric distribution also constrains that the ICM wind direction should not be close to the line-of-sight. At the downstream side, there are dust filaments mostly aligned with the ICM wind direction and large dust clouds being ablated by the ICM wind.
These results suggest a significant motion of the galaxy in the plane of sky. Even though the radial velocity of the galaxy is +820 km/s to the velocity of the cluster, for Abell~3627 likely in a major merger, the actual radial component of \eso{}'s motion relative to the surrounding ICM can be very different. 
    
\item Some UV bright, compact sources are discovered in the galaxy, mostly upstream of nearby dust clouds (Fig. \ref{fig:young_star_region} \& \ref{fig:young_star_color_mag}).
We suggest that they are young star clusters triggered by the compression of ram pressure, which offers evidence of RPS-triggered SF in RPS galaxies. On the other hand, the total SFR from these young star clusters is much smaller than the total SFR of the galaxy.

\item A large amount of the molecular gas (5.5 $\times 10^{9}$ M$_{\odot}$) has been revealed from the galaxy and the inner tail region (Fig. \ref{fig:apex_galaxy} \& \ref{fig:apex_tail}), corresponding to a molecular gas fraction of $\sim$ 16\%. On the other hand, the distribution of the molecular gas is asymmetric, with no significant CO emission detected from upstream and abundant CO emission downstream. CO emission is also detected in an inner tail region beyond the optical disk. This again shows the impact of RPS on the ISM distribution. \eso{} likely still retains most of its ISM so the stripping is probably still at the early stage, also aided by the slow development of stripping under the edge-on configuration.

\item Despite its long X-ray/H$\alpha$ tail and the detection of the molecular gas in the inner tail region, no SF is revealed in the tail region from the \hst\ data (Fig. \ref{fig:hist_scatter} \& \ref{fig:hist_scatter_2}), which is consistent with the lack of compact H$\alpha$ sources in the tail \citep{2010ApJ...708..946S,2013ApJ...777..122Z}. An upper limit of the SFE ($\sim$ 1.9 $\times$ 10$^{-12}$ yr$^{-1}$) is put on the inner tail region.

\item \eso{} is a galaxy undergoing RPS with long gaseous X-ray/H$\alpha$ tails. However, for the lack of SF downstream and the lack of asymmetry in the optical blue light, it would have been excluded from the ``jellyfish'' samples based on the ground optical data or UV survey data from e.g., {\em Galex}. Thus, \eso{} presents a great example to demonstrate the diversity of RPS galaxies and their observation signature. Samples based on different selection criteria should be combined for a more comprehensive understanding of RPS, their impact on galaxy evolution and the fate of the stripped ISM. 

\end{enumerate}
   
As demonstrated by this work and previous works \citep[e.g.][]{2007MNRAS.376..157C,2015AJ....150...59K,2016AJ....152...32A,2019ApJ...870...63C}, the \hst{} data are important for such kind of detailed analysis. More analysis with the \hst{} data and future wide-field survey data from {\em Euclid} and {\em Nancy Grace Roman Space Telescope} will allow us to understand the young stellar population and SFE in the RPS tails better, also with the CO data from e.g., {\em ALMA} and optical spectroscopic data from e.g., {\em MUSE}.

%% file: acknowledge.tex
\section*{Acknowledgements}

We thank Alessandro Boselli and Matteo Fossati for useful comments and discussion.
We thank the referee for helpful comments.
Support for this work was provided by the National Aeronautics and Space Administration through {\em Chandra} Award Number GO2-13102A and GO6-17111X issued by the {\em Chandra} X-ray Center, which is operated by the Smithsonian Astrophysical Observatory for and on behalf of the National Aeronautics Space Administration under contract NAS8-03060. Support for this work was also provided by the NASA grants HST-GO-12372.09, HST-GO-12756.08-A, 80NSSC18K0606 and the NSF grant 1714764.
This publication is based on data acquired with the Atacama Pathfinder Experiment (APEX) under programmes ID 88.B-0934(A) and 94.B-0766(A). APEX is a collaboration between the Max-Planck-Institut fur Radioastronomie, the European Southern Observatory, and the Onsala Space Observatory.
MC acknowledges support from STScI (HST-AR-14556.001-A), NSF (1910687), and NASA (19-ATP19-0188).
PJ acknowledges support from the project LM2018106 of the Ministry of Education, Youth and Sports of the Czech Republic and from the project RVO:67985815.

\section*{DATA AVAILABILITY}

The \hst{} raw data used in this paper are available to download at the The Barbara A. Mikulski Archive for Space Telescopes\footnote{https://archive.stsci.edu/hst/}. The \apex{} raw data are available to download at the ESO Science Archive Facility\footnote{http://archive.eso.org/cms.html}.
The reduced data underlying this paper will be shared on reasonable requests to the corresponding author.

%% file: color_excess.tex
\section{Blue sources in the tail region}\label{app:color_excess}
\begin{figure} 
    \includegraphics[width=0.47\textwidth]{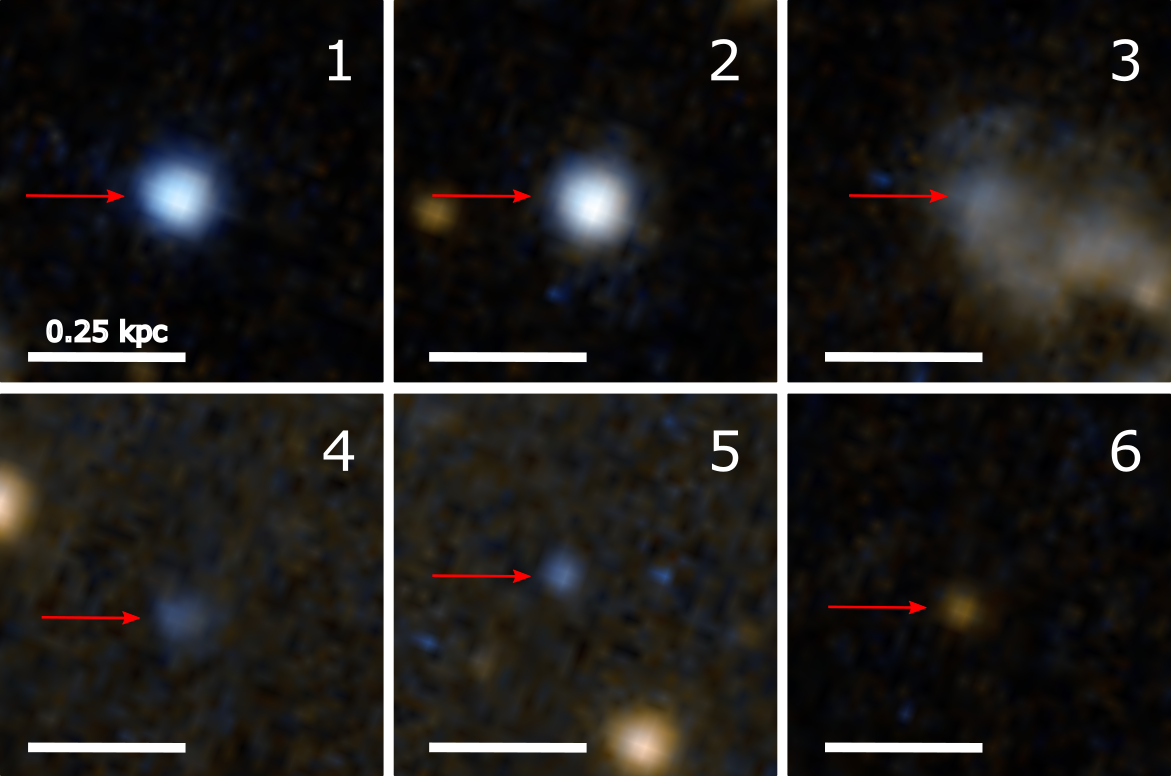}
    \vspace{-0.2cm}
    \caption{The composite image of the six sources in the tail region (zoom-in of Fig.~\ref{fig:composite}) with F275W $-$ F475W $\leq$ 2 and F475W $-$ F814W $\leq$ 1. Sources are numbered in order of the F475W flux. The brightest two objects are likely Galactic stars with significant F160W detections. Source \#3 is likely a background object.
    }
    \label{fig:tail_excess}
\end{figure}

There are six blue sources in the tail region as shown in Fig.~\ref{fig:tail_excess}. They are \#1 at (16:13:39.70, -60:53:21.17), \#2 at (16:13:40.56, -60:53:02.66), \#3 at (16:13:42.43, -60:53:36.55), \#4 at (16:13:37.55, -60:52:33.05), \#5 at (16:13:36.59, -60:52:31.57) and \#6 at (16:13:38.44, -60:52:51.09). The brightest two sources, \#1 and \#2, are likely Galactic stars with significant F160W emission and no H$\alpha$ emission. The morphology of the source \#3 suggests it is a background galaxy. If we exclude the source \#1 (also see its position in Fig.~\ref{fig:color_mag}), the total F275W flux of these five sources is 1.49 e$^{-}$/s, while the total F275W flux of sources in the control region selected with the same color cuts is 11.57 e$^{-}$/s. After correcting the difference on the sky area, the net F275W flux in the tail region is -0.25 e$^{-}$/s, again consistent with no recent SF present in the tail.

